\documentclass[%
 reprint,
 amsmath,
 amssymb,
 aps,
 pra,
 showpacs
]{revtex4-1}

\usepackage{graphicx}
\usepackage{dcolumn}
\usepackage{bm}
\usepackage{bbold}
\usepackage[export]{adjustbox}
\usepackage{amsmath,amsfonts,mathtools}
\usepackage{dsfont}
\usepackage{amsmath}
\usepackage{amssymb}
\usepackage{makeidx}
\usepackage{graphicx}
\usepackage{color}

\newcommand{\wrt}{w.r.t.\ }

\newcommand{\hide}[1]{}

\newcommand{\im}{\Im m\,}

\newcommand{\om}{\omega}

\newcommand{\la}{\lambda}

\newcommand{\lcase}{\left\{\begin{array}{ll}}
\newcommand{\rcase}{\end{array}\right.}
\newcommand{\ear}{\end{array}}
\newcommand{\bal}{\begin{align}}
\newcommand{\eal}{\end{align}}
\newcommand{\bma}{\begin{pmatrix}}
\newcommand{\ema}{\end{pmatrix}}
\newcommand{\beq}{\begin{equation}}
\newcommand{\eeq}{\end{equation}}
\newcommand{\bel}[1]{\begin{equation}\label{eq:#1}}
\newcommand{\eel}{\end{equation}}
\newcommand{\bea}{\begin{eqnarray}}
\newcommand{\eea}{\end{eqnarray}}
\newcommand{\beaNN}{\begin{eqnarray*}}
\newcommand{\eeaNN}{\end{eqnarray*}}

\newcounter{lecture}

\newcommand{\Ef}{\mathcal{E}}

\newcommand{\pde}{\partial}
\renewcommand{\hide}[1]{}

\newcommand{\rev}[1]{{ #1}}
\newcommand{\inten}[2]{#1\times10^{#2}\,\rev{\text{W}/\text{cm}^2}}
\newcommand{\optc}{\text{ opt.cyc.}}
\renewcommand{\im}{\text{i}}
\newcommand{\hydroplus}{\text{H}_2^+}

\begin{document}

\preprint{APS/123-QED}

\title{A quantum simulation of dissociative ionization of $H_2^+$ in full dimensionality with time-dependent surface flux method}
\author{Jinzhen Zhu}
\email{Jinzhen.Zhu@physik.uni-muenchen.de,zhujinzhenlmu@gmail.com}
\affiliation{%
 Physics Department, Ludwig Maximilians Universit\"at, D-80333 Munich, Germany
}%
\begin{abstract}
The dissociative ionization of $\hydroplus$ in a linearly polarized, 400 nm laser pulse is simulated by solving a three-particle time-dependent Schr\"odinger equation in full dimensionality without using any data from quantum chemistry computation.
The joint energy spectrum (JES) is computed using a time-dependent surface flux (tSurff) method, the details of which are given.
The calculated ground energy is -0.597 atomic units and internuclear distance is 1.997 atomic units if the kinetic energy term of protons is excluded, consistent with the reported precise values from quantum chemistry computation.
If the kinetic term of the protons is included, the ground energy is -0.592 atomic units with an internuclear distance 2.05 atomic units. 
Energy sharing is observed in JES and we find peak of the JES with respect to nuclear kinetic energy release (KER) is within $2\sim4$ eV, which is different from the previous two dimensional computations (over 10 eV), but is close to the reported experimental values.
The projected energy distribution on azimuth angles shows that the electron and the protons tend to dissociate in the direction of polarization of the laser pulse.
\end{abstract}

\pacs{32.80.-t,32.80.Rm,32.80.Fb}
\maketitle

\section{\label{sec:intro}Introduction}
Understanding the three-body Coulomb interaction problem is an on-going challenge in attosecond physics.
The typical candidates for investigation include Helium atom and $\hydroplus$ molecule.
In attosecond experiments, a short, intense laser pulse is introduced as a probe for the measurements.
Various mechanisms were proposed in the recent decades to describe the dissociation and dissociative ionization of $\hydroplus$, including bond softening~\cite{Bucksbaum1990},the charge-resonance enhanced ionization (CREI)~\cite{Zuo1995}, bond hardening~\cite{Yao1993}, above threshold dissociation (ATD)~\cite{Giusti-Suzor1995,Jolicard1992}, high-order-harmonic generation (HHG)~\cite{Zuo1993} and above threshold explosion~\cite{Esry2006}.
One may find a summary of the above mechanisms in theoretical and experimental investigations of $\hydroplus$ in literature~\cite{Posthumus2004,Giusti-Suzor1990}.
Experimental studies on the $\hydroplus$ ion exposed to circular and linearly polarized pulses for angular and energy distributions of electrons were reported recently~\cite{Odenweller2011,Odenweller2014,Wu2013,Gong2016}.
\par
In theory, the joint energy spectra (JES) of the kinetic energy release (KER) for one electron and two protons of the $\hydroplus$ ion are predominant observables that show how energy distributes around the fragments, where the JES is represented by the KER of two electrons for double ionization (DI)~\cite{Madsen2012,Scrinzi2012,Zielinski2016}.
In theory, the JES computations for double ionization in full dimensionality was very scarce for laser pulses with wavelengths beyond the XUV regime ($\geq 400$ nm)
because the computational consumption scales dramatically with the wavelength and intensity of the laser field~\cite{Zielinski2016}. 
With tSurff method, which was first introduced in Ref.~\cite{Tao2012}, full dimensional simulation of the JES for double ionization was available with moderate computational resources for 800 nm~\cite{Zielinski2016} and 400 nm~\cite{Zhu2020} laser pulses.
The tSurff method was also successfully applied to the dissociative ionization of the $\hydroplus$ ion~\cite{Yue2013,Yue2014} in a two-dimensional (2D) model, where the energy sharing of the photons and electron is observed in JES.
\par
The dissociative ionization of the $\hydroplus$ ion has been simulated by many groups~\cite{Steeg2003,Qu2002,Silva2013,Madsen2012,Odenweller2011,Takemoto2010,Feuerstein2003,Kulander1996}.
However, they are all in reduced dimensionality.
Quantum simulation in full dimensionality is not available yet.
Although the correlation among the fragments could be observed in the 2D model, the peaks of the JES with total nuclear KER are always above 10 eV.
This is far from experimental observables~\cite{Wu2013,Odenweller2014,Gong2016}, which are usually below 5 eV.
The tRecX code, which successfully implements the tSurff method in full dimensionality, has been applied successfully in the simulations of the double ionization of Helium~\cite{Zielinski2016} and the single ionization of polyelectron molecules~\cite{Majety2015,Majety2015e,Majety2015d,Majety2015c,Majety2015g}. 
The dissociative ionization of the $\hydroplus$ ion has not been computed using the tRecX code from before, even in reduced dimensionality.
\par
In this paper, we will introduce simulations of the dissociative ionization of the $\hydroplus$ ion by solving the time-dependent Schr\"odinger equation (TDSE) in full dimensionality based on the tRecX code.
We will first present the computational method for scattering amplitudes with tSurff methods, from which the JES can be obtained.
Then we will introduce the specific numerical recipes for the $\hydroplus$ ion based on the existing discretization methods of tRecX code.
With such numerical implementations, the {\it ab initio} calculation of field free ground energy of the Hamiltonian is available.
Finally we will present results of dissociative ionization in a 400 nm laser pulse, the JES, and projected energy spectrum on the azimuth angle.
\section{Methods}
In this paper, atomic units with specifying $\hbar=e^2=m_e=4\pi\epsilon_0\equiv1$ are used if not specified.
Center of the mass of two protons is chosen to be the origin.
Instead of using the vector between two protons $\vec{R}$ as an coordinate~\cite{Yue2013,Yue2014,Madsen2012}, we specify the coordinates of the protons and electrons as $\vec{r}_1,-\vec{r}_1$ and $\vec{r}_2$.
We denote $M=1836$ atomic units as the mass of the proton.
\subsection{Hamiltonian}
The total Hamiltonian can be represented by sum of the electron-proton interaction $H_{EP}$ and two tensor products, written as
\rev{
\begin{equation}\label{eq:HamiltonianH2PlusFull}
 H=H_{B}=H^{(+)}\otimes \mathds{1}+\mathds{1}\otimes H^{(-)}+H_{EP},
\end{equation}}
where the tensor products are formed by the identity operator $\mathds{1}$ multiplied by the Hamiltonian for two protons ($H^{(+)}$), or that for the electron ($H^{(-)}$).
$H_{B}$ is called the Hamiltonian in the $B$ region and will be detailed later.
With the coordinate transformation used in Ref.~\cite{Hiskes1961}, which is also illustrated in Appendix~\ref{sec:coordinates} for our specific case, the single operator for the electron is
\begin{equation}
 H^{(-)}=-\frac{\Delta}{2m}-\im\beta\vec{A}(t)\cdot\vec{\triangledown },
\end{equation}
and the Hamiltonian for protons can be written as
\begin{equation}
  H^{(+)}=-\frac{\Delta}{4M}+\frac{1}{2r},
\end{equation}
where we introduce reduced mass $m=\frac{2M}{2M+1}\approx1$ and $\beta=\frac{1+M}{M}\approx1$ for the electron, and $\vec{A}(t)$ is the vector potential.
The Hamiltonian of the electron-proton interaction can be written as
\rev{
\begin{equation}
 H_{EP}=-\frac{1}{|\vec{r}_1+\vec{r}_2|}-\frac{1}{|\vec{r}_1-\vec{r}_2|}.
\end{equation}}
\subsection{tSurff for dissociative ionization}
The tSurff method is applied here for the dissociative ionizations, which was successfully applied to the polyelectron molecules and to the double emission of He atom~\cite{Zielinski2016,Majety2015c,Majety2015d,Majety2015g,Majety2015e}.
In this section, we will follow a similar procedure as is done in Ref.~\cite{Zielinski2016}.
\par
According to the approximations of tSurff method, beyond a sufficient large tSurff radius $R_c^{(+/-)}$, the interactions of protons and electrons can be neglected, with the corresponding Hamiltonians being $H_V^{(+)}=-\frac{\Delta}{4M}$ for two protons and $H_V^{(-)}=-\frac{\Delta}{2m}-\im\beta\vec{A}(t)\cdot\vec{\triangledown}$ for the electron.
The scattered states of the two protons, which satisfy $\im\partial_t \chi_{\vec{k}_1}(\vec{r}_1)=H_V^{(+)}\chi_{\vec{k}_1}(\vec{r}_1)$, are
\begin{equation}
  \chi_{\vec{k}_1}(\vec{r}_1)=\frac{1}{(2\pi)^{3/2} }\exp(-\im\int_{t_0}^t\frac{k_1^2}{4M}d\tau)\rev{\exp(\im\vec{k}_1\cdot\vec{r}_1}),
\end{equation}
and those of the electron, which satisfies $\im\partial_t \chi_{\vec{k}_2}(\vec{r}_2)=H_V^{(-)}\chi_{\vec{k}_2}(\vec{r}_2)$, are
\begin{equation}
 \chi_{\vec{k}_2}(\vec{r}_2)=\frac{1}{(2\pi)^{3/2} }\exp(-\im\int_{t_0}^t \frac{k_2^2}{2m}-\im\beta\vec{A}(\tau)\cdot\vec{\triangledown}d\tau)\rev{\exp(\im\vec{k}_2\cdot\vec{r}_2)},
\end{equation}
where we assume the laser field starts at $t_0$ and $\vec{k}_{1/2}$ denote the momenta of the protons or the electron.
\par
\begin{figure}
\centering
\includegraphics[width=0.4\textwidth]{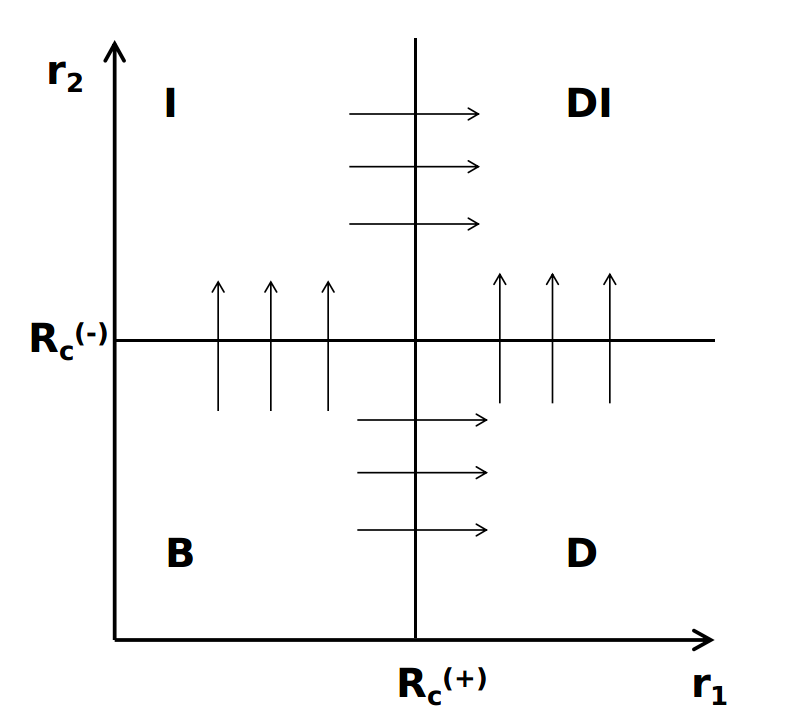}
\caption{The regions of dissociative ionization time propagation.
The B stands for bound region, D for dissociation region where the two protons are out of $R_c^{(+)}$ but electron not ionized and stays inside.
I represents the ionization region where electron is out-of-box $R_c^{(-)}$ but two protons are still inside $R_c^{(+)}$.
DI stands for the dissociative ionization region where both the electron and the protons are out of $R_c^{(+/-)}$. $R_c^{(+/-)}$ are the tSurff radii for $r_1=|\vec{r}_1|$ or $r_2=|\vec{r}_2|$.
}
\label{fig:H2PlusRegions}
\end{figure}

Based on the tSurff radius $R_c^{(+/-)}$, we may split the dissociative ionization into four regions namely $B,I,D,DI$, shown in figure~\ref{fig:H2PlusRegions}, where bound region $B$ preserves the full Hamiltonian in Eq.~(\ref{eq:HamiltonianH2PlusFull}), $D,I$ are time propagations by single particles with the Hamiltonian
\begin{equation}
 H_{D}(\vec{r}_2,t)=H_{V}^{(-)}(\vec{r}_2,t) = -\frac{\Delta}{2m}-\im\beta\vec{A}(t)\cdot\vec{\triangledown}
\end{equation}
and
\begin{equation}
 H_{I}(\vec{r}_1,t) = -\frac{\Delta}{4M}+\frac{1}{2r_1},
\end{equation}
and $DI$ is an integration process.
The treatment was first introduced in the double ionization of Helium in Ref.~\cite{Scrinzi2012} and then applied in a 2D simulation of the $\hydroplus$ ion in Ref.~\cite{Yue2013}.
Without considering the low-energy free electrons that stay inside the box after time propagation, we may write
\begin{equation}
 \begin{split}
  &\psi_{B}(\vec{r}_1,\vec{r}_2,t)\approx 0,\quad r_1\geq R_c^{(+)},\text{or}\;r_2\geq R_c^{(-)}\\
  &\psi_{D}(\vec{r}_1,\vec{r}_2,t)\approx 0,\quad r_1<R_c^{(+)},\text{or}\;r_2\geq R_c^{(-)}\\
  &\psi_{I}(\vec{r}_1,\vec{r}_2,t)\approx 0,\quad r_1\geq R_c^{(+)},\text{or}\;r_2<R_c^{(-)}\\
  &\psi_{DI}(\vec{r}_1,\vec{r}_2,t)\approx 0,\quad r_1<R_c^{(+)},\text{or}\;r_2<R_c^{(-)}\\
 \end{split}
\end{equation}
\par
We assume that for a sufficiently long propagation time $T$, the scattering ansatz of electron and protons disentangle.
By introducing the step function 
\begin{equation}
 \Theta_{1/2}(R_c)=\left\{\begin{matrix}
0\;, r_{1/2}< R_c^{(+/-)} \\ 
1\;, r_{1/2}\geq  R_c^{(+/-)},
\end{matrix}\right.
\end{equation}
the unbound spectra can be written as
\begin{equation}\label{eq:finalSpectrum}
 P(\vec{k}_1,\vec{k}_2)=P(\phi_1,\theta_1,k_1,\phi_2,\theta_2,k_2)=\left | b(\vec{k}_1,\vec{k}_2,T) \right |^2.
\end{equation}
$b(\vec{k}_1,\vec{k}_2,T)$ \rev{are} the scattering amplitudes and can be written as
\begin{equation}\label{eq:integralAmplitudes}
\begin{split}
b(\vec{k}_1,\vec{k}_2,T)=&\langle \chi_{\vec{k}_1}\otimes \chi_{\vec{k}_2} |\Theta_1(R_c)\Theta_2(R_c)|\psi(\vec{r}_1,\vec{r}_2,t)\rangle \\
=&\int_{-\infty}^{T}[F(\vec{k}_1,\vec{k}_2,t)+\bar{F}(\vec{k}_1,\vec{k}_2,t)] dt
\end{split}
\end{equation}
with two sources written as
\begin{equation}\label{eq:Fk1k2H2Plus}
 F(\vec{k}_1,\vec{k}_2,t)= \langle \chi_{\vec{k}_2}(\vec{r}_2,t)\left |[H_V^{(-)}(\vec{r}_2,t), \Theta_2(R_c)] \right | \varphi_{\vec{k}_1}(\vec{r}_2,t) \rangle
\end{equation}
and 
\begin{equation}\label{eq:Fk1k2H2PlusBar}
 \bar{F}(\vec{k}_1,\vec{k}_2,t)= \langle \chi_{\vec{k}_1}(\vec{r}_1,t)\left | [H_V^{(+)}(\vec{r}_1,t), \Theta_1(R_c)] \right | \varphi_{\vec{k}_2}(\vec{r}_1,t) \rangle.
\end{equation}
The single particle wavefunctions $\varphi_{\vec{k}_1}(\vec{r}_2,t)$ and $\varphi_{\vec{k}_2}(\vec{r}_1,t)$ satisfy
\begin{equation}\label{eq:unbound0}
 \im\frac{d}{dt}\varphi_{\vec{k}_1 }(\vec{r}_2,t)=H_D(\vec{r}_2,t) \varphi_{\vec{k}_1 }(\vec{r}_2,t)-C_{\vec{k}_1 }(\vec{r}_2,t)
\end{equation}
and
\begin{equation}\label{eq:unbound1}
 \im\frac{d}{dt}\varphi_{\vec{k}_2 }(\vec{r}_1,t)=H_I(\vec{r}_1,t)\varphi_{\vec{k}_2 }(\vec{r}_1,t)-C_{\vec{k}_2 }(\vec{r}_1,t).
\end{equation}
The sources are the overlaps of the two-electron wavefunction \rev{and} the Volkov solutions shown by 
\begin{equation}\label{eq:source1}
 C_{\vec{k}_1 }(\vec{r}_2,t)=\int d\vec{r}_1 \overline{\chi_{\vec{k}_1}(\vec{r}_1,t)}[H_V^{(+)}(\vec{r}_1,t),\Theta_1(R_c)]\psi(\vec{r}_1,\vec{r}_2,t)
\end{equation}
and
\begin{equation}\label{eq:source2}
 C_{\vec{k}_2 }(\vec{r}_1,t)=\int d\vec{r}_2 \overline{\chi_{\vec{k}_2}(\vec{r}_2,t)}[H_V^{(-)}(\vec{r}_2,t),\Theta_2(R_c)]\psi(\vec{r}_1,\vec{r}_2,t),
\end{equation}
with initial values being 0, where $\overline{\cdots}$ means complex conjugate.
The two tSurff radii could be set to equivalent $R_c^{(+)}=R_c^{(-)}$, because all Coulomb interactions are neglected when either the protons or electron is out of the tSurff radius.
According to our previous researches, the spectrum computation is independent of the $R_c$ if all Coulomb terms are removed and the wavefunction is propagated long enough after the pulse~\cite{Zielinski2016,Scrinzi2012}.
The tSurff for double emission of two particles was firstly introduced in Ref.~\cite{Scrinzi2012}.
The above derivations are very similar to what was reported in Ref.~\cite{Zielinski2016} of double emission of Helium, where the only differences are constants before different operators, say $\Delta, \vec{\triangledown}$ and $\frac{1}{r}$.
Thus, detailed formulas are omitted here and the interested readers can refer to Ref.~\cite{Zielinski2016,Scrinzi2012}.
\par
The computation for photoelectron spectrum includes four steps, similar to the one used in Ref.~\cite{Zielinski2016}, detailed as
\begin{enumerate}
 \item Solve \rev{the} full 6D TDSE with the Hamiltonian in the $B$ region, given in Eq.~(\ref{eq:HamiltonianH2PlusFull}), and write the time-dependent surface values in the disk.
 \item Evolve the single-particle wave packets in the $D$ region by Eq.~(\ref{eq:Fk1k2H2Plus}) with surface values given in the $B$ region time propagation.
 \item Evolve the single-particle wave packets in $I$ region by Eq.~(\ref{eq:Fk1k2H2PlusBar}) with the surface values given in the $B$ region time propagation.
 \item Integrate the fluxes calculated from surface values written in the $D$ and $I$ regions' time propagation by Eq.~(\ref{eq:integralAmplitudes}).
\end{enumerate}
\section{Numerical implementations}
The numerical methods here are similar to what was detailed in Ref.~\cite{Zielinski2016,Zhu2020}.
In fact, the code in this paper is developed based on the double ionization framework of the tRecX code used in the reference~\cite{Zielinski2016,Zhu2020}.
Thus, we will focus on the electron-protons interaction which was not mentioned before and only list relevant discretization methods in this paper. 
\subsection{Discretization and basis functions}
The 6D wavefunction $\psi$ is represented by the product of spherical harmonics for angular momentum and radial functions as
\begin{equation}\label{eq:DIWf}
\begin{split}
 &\psi(\vec{r}_1,\vec{r}_2,t)=\psi(r_1,\theta_1,\phi_1,r_2,\theta_2,\phi_2,t)\\
 =&\sum_{m_1,l_1,m_2,l_2}Y^{m_1}_{l_1}(\theta_1,\phi_1)Y^{m_2}_{l_2}(\theta_2,\phi_2)R_{m_1,m_2,l_1,l_2}(r_1,r_2,t),
\end{split}
\end{equation}
where $Y^{m_1}_{l_1}(\theta_1,\phi_1)$ and $Y^{m_2}_{l_2}(\theta_2,\phi_2)$ are the spherical harmonics of the two electrons and the radial function is represented by the finite-element discrete variable representation (FE-DVR) method as
\begin{equation}
\begin{split}
 R_{m_1,m_2,l_1,l_2}(r_1,r_2,t)=&\sum_{n_1,n_2} R_{m_1,m_2,l_1,l_2}^{n_1,n_2}(r_1,r_2,t)\\
 R_{m_1,m_2,l_1,l_2}^{n_1,n_2}(r_1,r_2,t)=&\sum_{p_1,p_2}f_{p_1}^{(n_1)}(r_1)f_{p_2}^{(n_2)}(r_2)\frac{1}{r_1r_2}\\
				& c_{n_1,n_2,p_1,p_2}^{m_1,m_2,l_1,l_2}(t)
\end{split}
\end{equation}
where $f_{p_{1/2} }^{(n_{1/2})}(r_{1/2})$ are $p_{1/2}$th basis functions on $n_{1/2}$th element,
and the time-dependency of the three particles are included in the radial functions and coefficients $c_{n_1,n_2,p_1,p_2}^{m_1,m_2,l_1,l_2}(t)$, as is used in Ref.~\cite{Zielinski2016, Scrinzi2012}.
The infinite-range exterior complex scaling (irECS) method is utilized as an absorber~\cite{Scrinzi2010}.
The tSurff expression for computing spectra of such discretization can be found in Ref.~\cite{Zielinski2016}.
\subsection{Electron-protons interaction}
The first part of electron-protons interaction can be written in a multi-pole expansion as
\begin{equation}
\begin{split}
  \frac{1}{|\vec{r}_1-\vec{r}_2|}=&\frac{1}{\sqrt{r_1^2+r_2^2-2r_1r_2\cos\gamma}}\\
  =&\frac{1}{r_{>}}\frac{1}{\sqrt{1+h^2-2h\cos\gamma}}=\sum_{l=0}^{\infty}\frac{h^l}{r_{>}}P_l(\cos\gamma),
\end{split}
\end{equation}
where $r_>=\max(r_1,r_2),r_<=\min(r_1,r_2), h=\frac{r_<}{r_>}$, $\gamma$ is the angle between $\vec{r}_1,\vec{r}_2$ and $P_l(\cos\gamma)$ are Legendre polynomials.
Similarly, we have
\begin{equation}
  \frac{1}{|\vec{r}_1+\vec{r}_2|}=\frac{1}{r_{>}}\frac{1}{\sqrt{1+h^2+2h\cos\gamma}}=\sum_{l=0}^{\infty} (-1)^l \frac{h^l}{r_{>}} P_l(\cos\gamma).
\end{equation}
And the summation goes as
\begin{equation}
  \frac{1}{|\vec{r}_1+\vec{r}_2|}+\frac{1}{|\vec{r}_1-\vec{r}_2|}=2\sum_{l=0}^{\infty} \frac{h^l}{r_{>}}P_l(\cos\gamma)\quad l\bmod 2=0.
\end{equation}
where $l \bmod 2=0$ means $l$ is even.
With the Legendre polynomials expanded by spherical harmonics $Y_l^{m}(\theta_2,\phi_2)$ and $Y_l^{m*}(\theta_1,\phi_1)$ , we have
\rev{
\begin{equation}
\begin{split}
  H_{EP}&=-2\sum_{l=0}^{\infty}\sum_{m=-l}^{l} \frac{4\pi}{2l+1}\frac{r_<^{l} }{r_>^{l+1}}Y_l^{m}(\theta_2,\phi_2) Y_l^{m*}(\theta_1,\phi_1)\\
  &\quad l\bmod 2=0.
\end{split}
\end{equation}
}The matrix elements of electron-protons are
\rev{
\begin{equation}\label{eq:EP}
\begin{split}
  &\langle\psi^{(n_1',n_2')}_{m_1',m_2',l_1',l_2'}|-\frac{1}{|\vec{r}_1-\vec{r}_2|}-\frac{1}{|\vec{r}_1+\vec{r}_2|}|\psi^{(n_1,n_2)}_{m_1,m_2,l_1,l_2}\rangle\\
=&-2\sum_{\lambda\mu}\frac{4\pi}{2\lambda+1}\langle Y_{l_1'}^{m_1'}Y_{\lambda}^{\mu}|Y_{l_1}^{m_1}\rangle\langle Y_{l_2'}^{m_2'}|Y_{\lambda}^{\mu}Y_{l_2}^{m_2}\rangle\\
&\langle R^{n_1',n_2'}_{m_1',m_2',l_1',l_2'}|\frac{r_<^\lambda}{r_>^{\lambda+1}}|R^{n_1,n_2}_{m_1,m_2,l_1,l_2}\rangle,\lambda \bmod 2=0,
\end{split}
\end{equation}
}which could be obtained by dropping the odd $\lambda$ terms and \rev{multiplying the} even $\lambda$ terms by \rev{-2} in the standard multi-pole expansion for electron electron interactions from Ref.~\cite{Zielinski2016} as
\begin{equation}
\begin{split}
  &\langle\psi^{(n_1'n_2')}_{m_1'm_2'l_1'l_2'}|\frac{1}{|\vec{r}_1-\vec{r}_2|}|\psi^{(n_1n_2)}_{m_1,m_2,l_1,l_2}\rangle\\
=&\sum_{\lambda\mu}\frac{4\pi}{2\lambda+1}\langle Y_{l_1'}^{m_1'}Y_{\lambda}^{\mu}|Y_{l_1}^{m_1}\rangle\langle Y_{l_2'}^{m_2'}|Y_{\lambda}^{\mu}Y_{l_2}^{m_2}\rangle\\
&\langle R^{n_1'n_2'}_{m_1',m_2',l_1',l_2'}|\frac{r_<^\lambda}{r_>^{\lambda+1}}|R^{n_1n_2}_{m_1,l_1,m_2,l_2}\rangle.
\end{split}
\end{equation}
Here 
\begin{equation}
 \psi^{(n_1,n_2)}_{m_1,m_2,l_1,l_2}=Y^{m_1}_{l_1}(\theta_1,\phi_1)Y^{m_1}_{l_1}(\theta_2,\phi_2)R_{m_1,m_2,l_1,l_2}^{n_1,n_2}(r_1,r_2,t).
\end{equation}
Therein, the matrix for electron-protons interaction could be obtained by the numerical recipes used in Ref.~\cite{Zielinski2016,McCurdy2004} with limited changes.
Numerically, we find $\lambda$ does not need to go to infinity and a maximum value of 8 already suffices our simulations.
\section{Numerical results}
A numerical convergence study shows, unlike the 6D double emission of $\text{He}$, where $m_{1/2}=0,\quad 0\leq l_{1/2}\leq 2$ already gives convergent ground eigenenergy~\cite{Zielinski2016}, here the angular quantum number $0\leq m_{1/2} \leq 2$ and $0\leq l_{1/2}\leq 8$ starts to give convergent calculations, due to the lower symmetric property of the $\hydroplus$ ion.
The $R_c^{(+)}=R_c^{(-)}=12.5$ atomic units is chosen for computation, as we find $R_c^{(-)}$ does not change the quality of the spectrum but introduces longer propagation time for low-energy particles to fly out.
$R_c^{(+)}=12.5$ atomic units gives the internuclear distance $R=25$ atomic units as used in Ref.~\cite{Yue2013}.
According to the convergence study in appendix~\ref{sec:convergence}, $R_c^{(+)}=R_c^{(-)}=12.5$ atomic units gives JES with error below 10\%.
$R_c^{(-)}=12.5$ atomic units is much larger than the quiver radius of electron in a 400 nm, $\inten{8.3}{13}$ laser pulse.
Ref.~\cite{Chelkowski1995} shows the Coulomb explosion at large distances contributes to low energy fragments of protons, which are highly correlated to the resonances between two $\hydroplus$ eigenstates.
Thus we need a large simulation box to include the all possible eigenenergies.
In Ref.~\cite{Chelkowski1995}, the maximum internuclear distance for the low energy fragments is 11 atomic units.
The molecular eigenenergies are nearly invariant with internuclear distance far below our $2R_c^{(+)}=25$ atomic units here.
Any potential dynamic dissociation quenching effect (DDQ) is also included in $B$ region \rev{because} the $\hydroplus$ is in dissociative limit with internuclear distance over 12 atomic units~\cite{Chateauneuf1998}.
The wavefunction is propagated long enough after the pulse to include the unbound states \rev{with} low kinetic energies.
\par
If the kinetic energy of protons is included, the field free ground energy value is $E_0=-0.592$ atomic units and the internuclear distance is 2.05 atomic units.
With the kinetic energy of protons excluded, the ground eigenenergy is -0.597 atomic units, three digits exact to ground energy from quantum chemistry calculations in Ref.~\cite{Bressanini1997}, where the internuclear distance is fixed.
The internuclear distance is 1.997 atomic units, three digits exact to that from the precise computations in Ref.~\cite{Schaad1970}.
\subsection{Laser pulses}
\label{sec:pulses}
The dipole field of a laser pulse with peak intensity $I=\Ef_0^2$ (atomic units) and linear polarization in $z$-direction is defined as $\Ef_z(t)=-\pde_tA_z(t)$, phase $\phi_{CEP}=0$ with
\beq
A_z(t)=\frac{\Ef_0}{\om} a(t)\sin(\om t+\phi_{CEP}).
\eeq
A pulse with $\la=400$ \rev{nm} is given with intensities $\inten{8.3}{13}$ close to 2D computation in Ref.~\cite{Yue2013} and $\inten{5.9}{13}$ close to experimental conditions in Ref.~\cite{Wu2013}.
We choose $a(t)=[\cos(t/T)]^8$ as a realistic envelope. Pulse durations are specified as FWHM=5 $\optc$ \wrt intensity.
To compare with the published results, a 400 nm, $\sin^2$ envelope laser pulse at $\inten{8.8}{13}$ from Ref.~\cite{Yue2013} and a 791 nm, $\cos^8$ laser pulse at $\inten{7.7}{13}$ used in Ref.~\cite{Pavicic2005} are also applied.
\subsection{Joint energy spectra}
\begin{figure}
\centering
\includegraphics[width=0.23\textwidth,trim=0.1cm 0.1cm 0.1cm 0.1cm,clip]{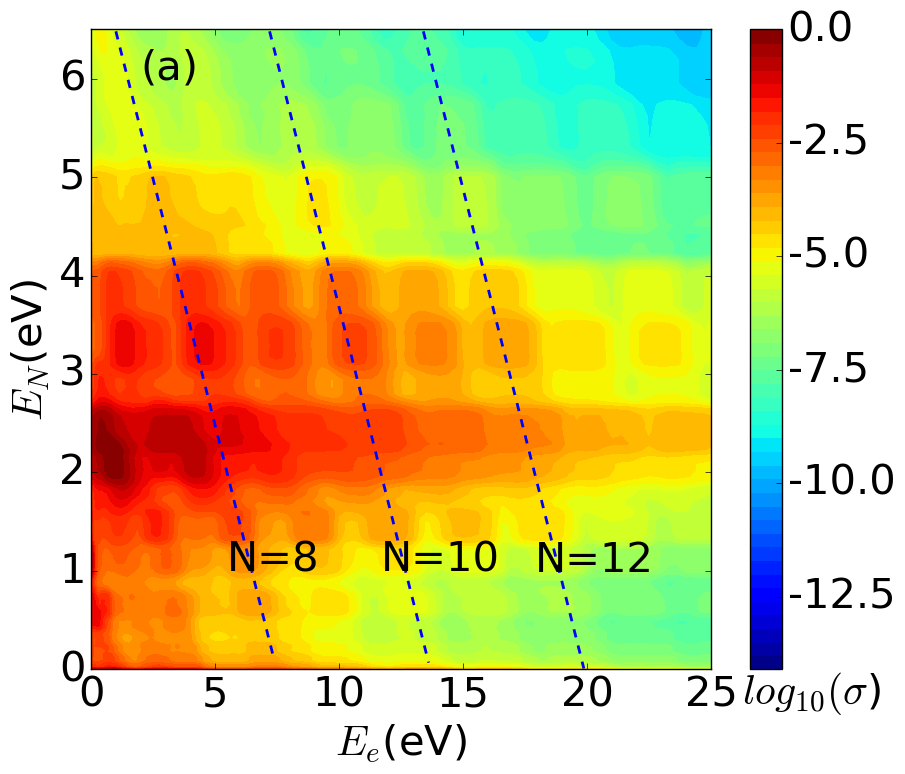}
\includegraphics[width=0.23\textwidth,trim=0.1cm 0.1cm 0.1cm 0.1cm,clip]{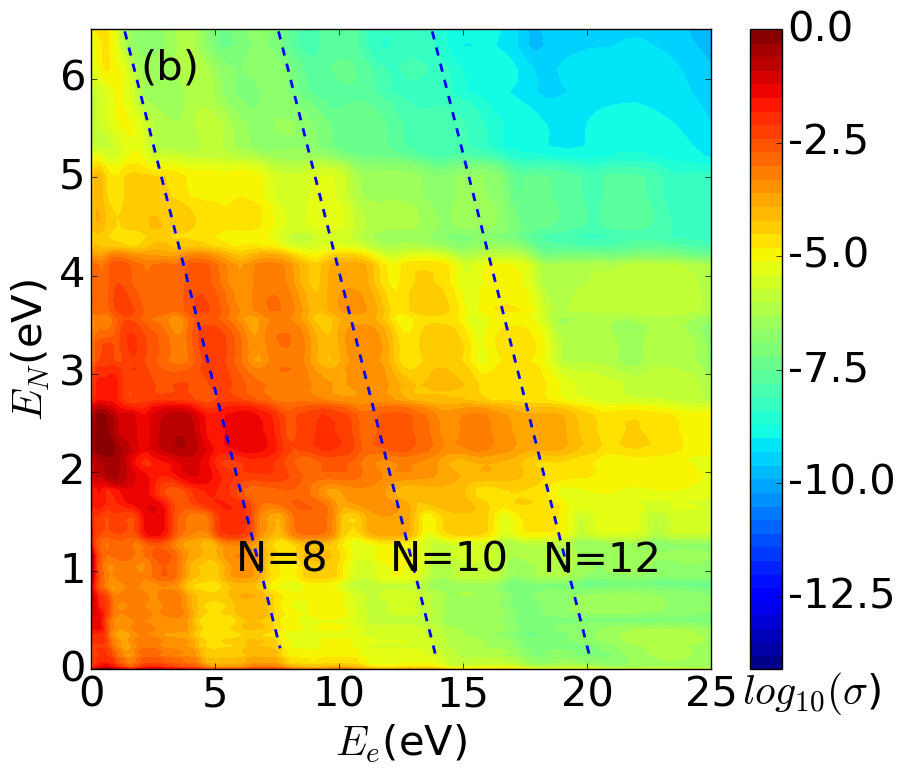}
\includegraphics[width=0.23\textwidth,trim=0.1cm 0.1cm 0.1cm 0.1cm,clip]{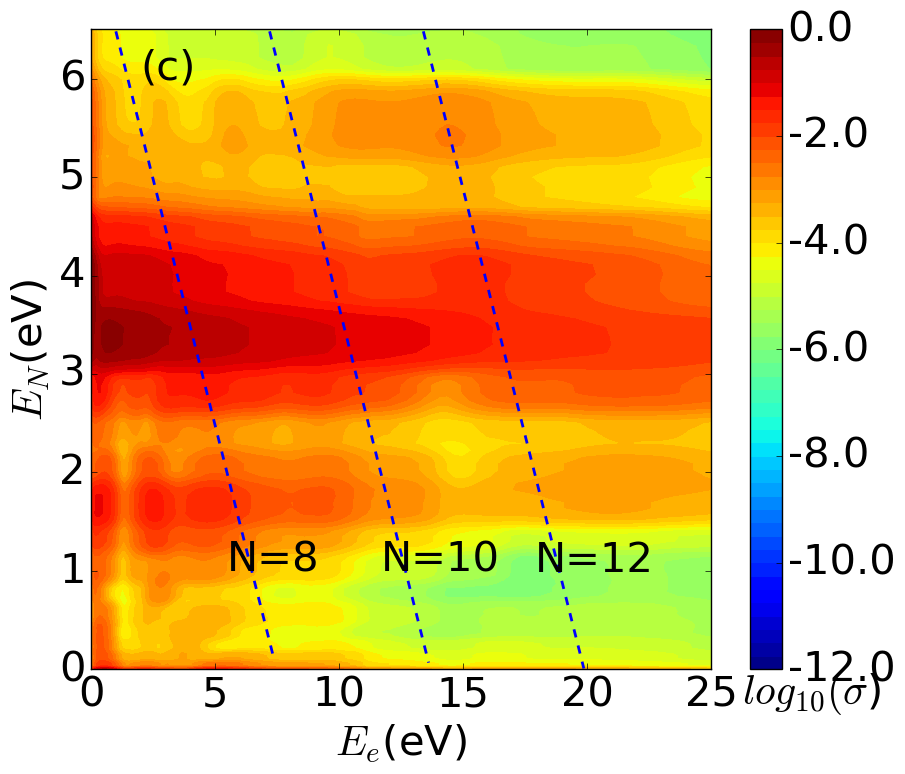}
\includegraphics[width=0.23\textwidth,trim=0.1cm 0.1cm 0.1cm 0.1cm,clip]{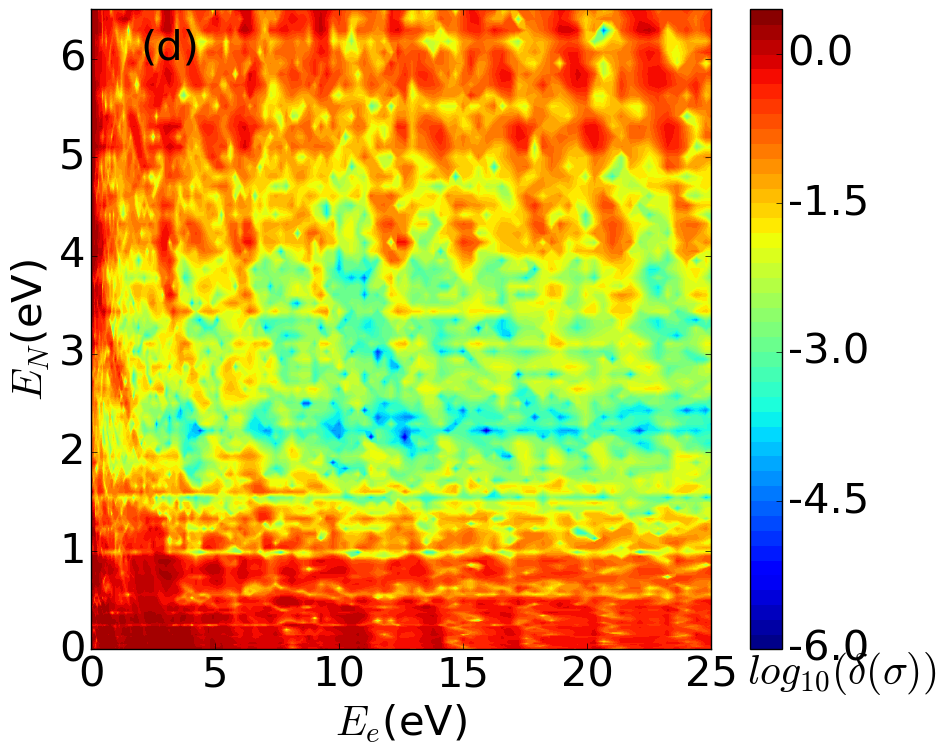}
\caption{Log-scale JES $\log_{10}\sigma(E_{N},E_{e})$ represented by total energy of two protons $E_{N}$ and that of an electron $E_{e}$.
Linear polarized, 400 nm, with (a) $\cos^8$ envelope with FWHM=5 $\optc$ at $\inten{8.3}{13}$ and (b) $\cos^8$ envelope with FWHM=5 $\optc$ pulses at $\inten{5.9}{13}$ is applied to the $\hydroplus$ ion. The dashed lines represent the energy sharing between the protons and electron with formula $E_{N}+E_{e}=N\omega+E_0-U_p$, where $\omega$ is the photon energy.
(c) JES from Coulomb explosion simulation from the ground eigenstate of the $\hydroplus$ ion. 
(d) Log-scale error $\log(\delta(\sigma))$ of two spectra from $\cos^8$ envelope laser pulse at $\inten{8.3}{13}$ with and without the contribution from $D\to DI$ (from Eq.~(\ref{eq:Fk1k2H2Plus})) by $\delta(\sigma)=2\frac{|\sigma'(E_{N},E_{e}) - \sigma(E_{N},E_{e})|}{|\sigma'(E_{N},E_{e}) + \sigma(E_{N},E_{e})|}$.
$\sigma(E_{N},E_{e})$ of (a) and (b) are normalized \rev{with dividing by} the maximum value.
}
\label{fig:H2PlusJES}
\end{figure}

The JES of the two dissociative protons and the electron is obtained by integrating Eq.~(\ref{eq:finalSpectrum}) over angular coordinates as
\rev{
\begin{equation}
\begin{split}
 \sigma(E_N,E_e)=&\int d\phi_1\int d\phi_2\int d\theta_1\sin\theta_1\int d\theta_2\sin\theta_2\\
 &P(\phi_1,\theta_1,\sqrt{4M E_N},\phi_2,\theta_2,\sqrt{2m E_{e} }),
 \end{split}
\end{equation}
}
where $E_N,E_e$ are kinetic energies of {\it two protons} and an electron, respectively.
$\sigma(E_{N},E_{e})$ is presented in Fig.~\ref{fig:H2PlusJES} (a, b).
The tilt lines with formula $E_{N}+E_{e}=N\omega+E_0 -U_p$ with ponderomotive energy $U_p=\frac{A_0^2}{4m}$ specify the energy sharing of $N$ photons for both the computations from $\inten{8.3}{13}$ and $\inten{5.9}{13}$, indicating correlated emissions of the electron and protons, which is also observed in the 2D computations~\cite{Yue2013,Yue2014}.
The yields are intense around nuclear KER from 2 eV to 4 eV in the $\cos^8$ envelope pulse, consistent with the experimental values reported in Ref.~\cite{Gong2016,Wu2013}.
The peak of JES for dissociative ionization is for lower nuclear KER than that (3-4 eV) from Coulomb explosion from ground eigenstate of the $\hydroplus$ ion, which property is also close to experimental observables~\cite{Pavicic2005}.
The Coulomb explosion JES is obtained with the same method as dissociative ionization except that $H_{EP}$ is removed from $B$ region Hamiltonian as $H^{(CS)}_{B}=H^{(+)}\otimes \mathds{1}+\mathds{1}\otimes H^{(-)}$, but the initial state is still obtained from Hamiltonian $H_B$ in Eq.~(\ref{eq:HamiltonianH2PlusFull}).
We find that the contribution from time-propagation in sub-region $D\to DI$ (see Eq.~(\ref{eq:Fk1k2H2Plus})) is small, as the numerical error of JES $\delta(\sigma)$ of $\sigma$ computed from $I\to DI$, and $\sigma'$ computed from two subregions ($I\to DI$ and $D\to DI$), is always below 1\% the main contribution of the JES ($2<E_{N}<4$ \rev{eV}), see Fig.~\ref{fig:H2PlusJES} (d).
This numerical property is also observed in two-dimensional (2D) simulations~\cite{Yue2013}.
This is because the electrons are much faster than protons and the $\hydroplus$ ion tends to release first.
\begin{figure}
\centering
\includegraphics[width=0.23\textwidth,trim=0.1cm 0.1cm 0.1cm 0.1cm,clip]{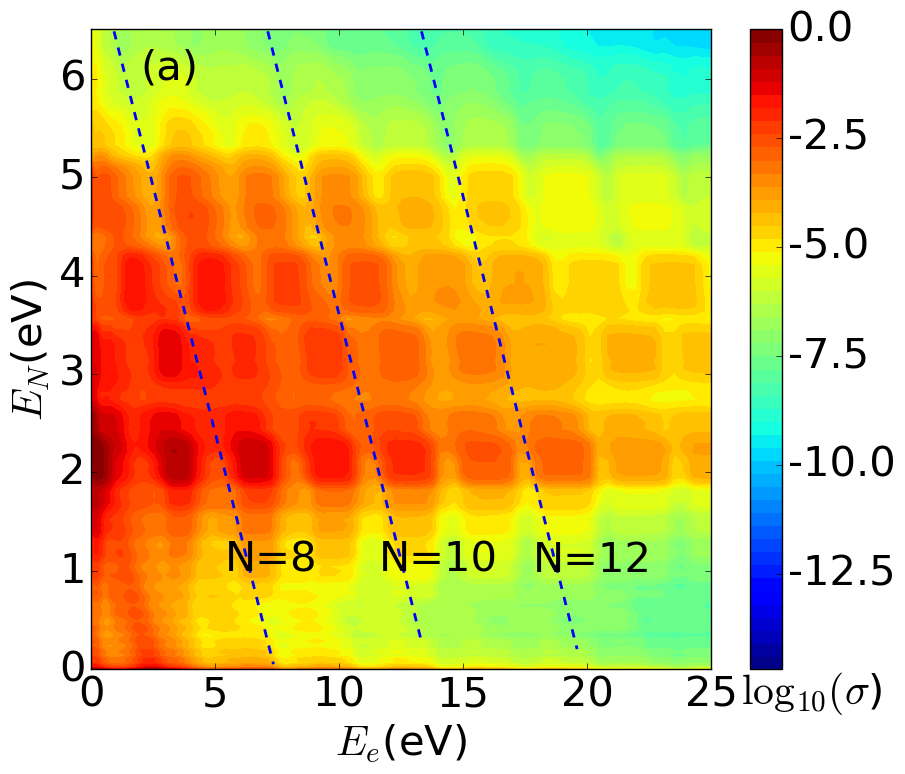}
\includegraphics[width=0.23\textwidth,trim=0.1cm 0.1cm 0.1cm 0.1cm,clip]{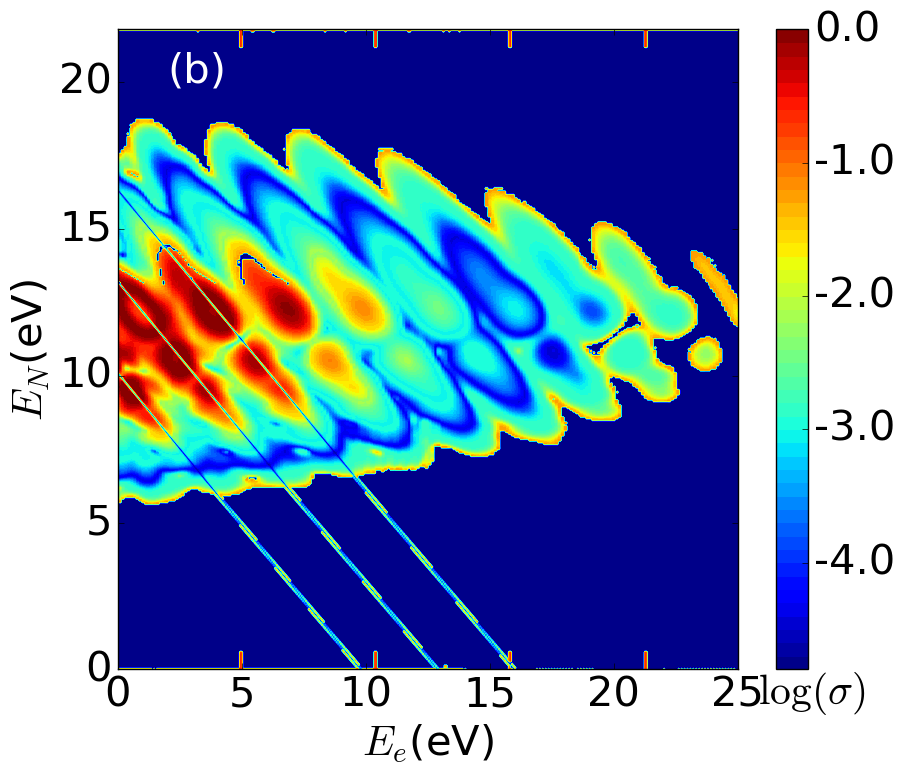}
\includegraphics[width=0.23\textwidth,trim=0.1cm 0.1cm 0.1cm 0.1cm,clip]{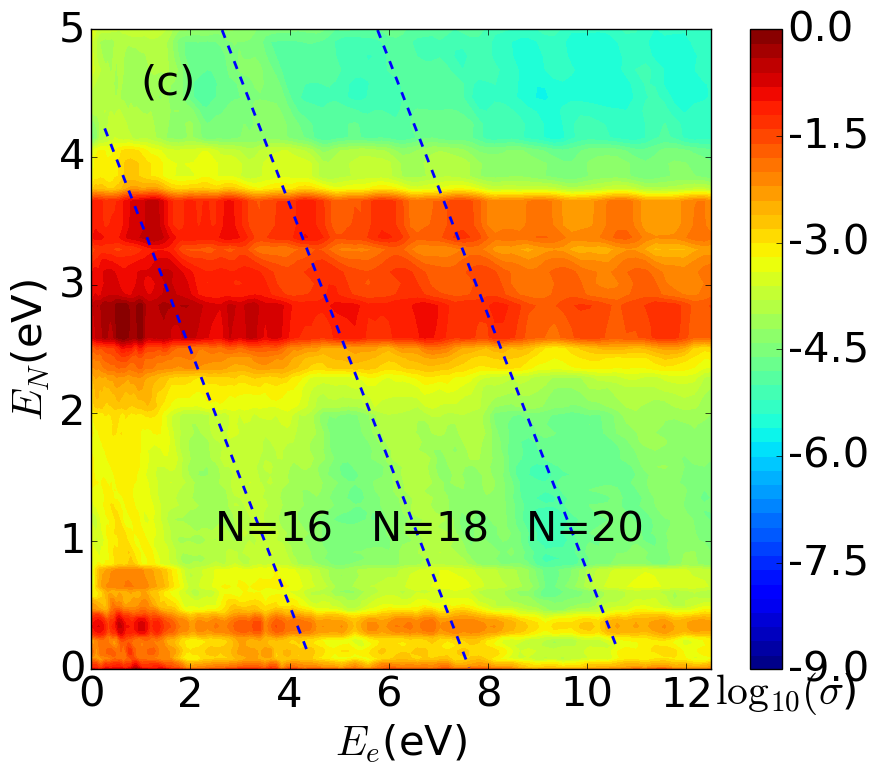}
\includegraphics[width=0.23\textwidth,trim=0.1cm 0.1cm 0.1cm 0.1cm,clip]{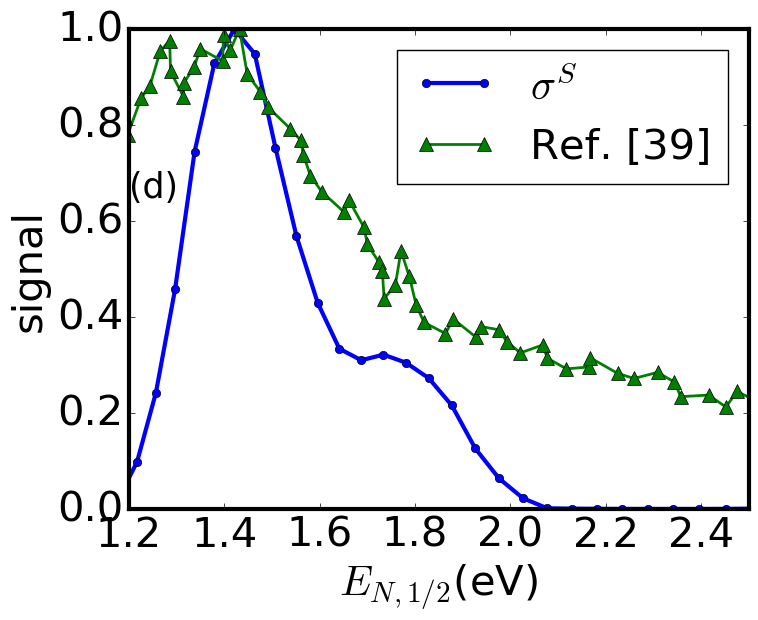}
\caption{Log-scale JES $\log_{10}\sigma(E_{N},E_{e})$ represented by total energy of two protons $E_{N}$ and that of an electron $E_{e}$.
JES of $\hydroplus$ in a linearly polarized, 400 nm laser pulse at $\inten{8.8}{13}$ as is used in Ref.~\cite{Yue2013} by (a) 6D computation compared to (b) the normalized 2D data from Ref.~\cite{Yue2013}, and (c) in a linearly polarized, 791 nm laser pulse at $\inten{7.7}{13}$ as used in Ref.~\cite{Pavicic2005}.
(d) Blue dots: The normalized single JES is created from $\sigma^S(E_{N,1/2})=\int \sigma(2E_{N,1/2},E_{e}) \sqrt{E_e} dE_e$, where $E_{N,1/2}$ depicts the kinetic energy of a proton. Green triangles: The normalized signals extracted from Ref.~\cite{Pavicic2005} with respect to the kinetic energy of a proton.
}
\label{fig:compareRef}
\end{figure}
\par
Here we compare our computations with the published results.
First, the JES with a 400 nm, $\sin^2$ envelope laser pulse at $\inten{8.8}{13}$ as used in the 2D simulations in Ref.~\cite{Yue2013} is computed, see Fig.~\ref{fig:compareRef} (a), compared to the results from 2D simulations in Fig.~\ref{fig:compareRef} (b). One clearly sees the JES is most considerable with nuclear KER around 2-4 eV in our computation but is with nuclear KER above 10 eV in the 2D simulation.
We also attach the JES with a linearly polarized, 791 nm laser pulse at $\inten{7.7}{13}$ as used in Ref.~\cite{Pavicic2005}, where JES is \rev{most} considerable with nuclear KER around 3 eV.
For the 791 nm computation, $R^{(-)}_c=15$ atomic units is applied which is slightly above the quiver radius of the electron.
For a direct comparison, we integrate the JES \rev{over} the electron KER and obtain the photoelectron spectrum with respect to the nuclear KER in Fig.~\ref{fig:compareRef} (d), where the experimental data from Ref.~\cite{Pavicic2005} is also attached.
The peak of the spectrum around 1.42 eV in our computation is in good agreement with the experimental data, and the position of a minor peak around 1.7 eV in our computation also matches the experimental observation.
The observation that JES is most considerable with nuclear KER around 2-4 eV can also be found in the Coulomb explosion computation shown in Fig.~\ref{fig:H2PlusJES} (c).
In other experiments, the distribution of emitted protons peaks at nuclear KER=4 eV for a 780 nm laser pulse at $\inten{6}{14}$~\cite{Odenweller2014}, Ref.~\cite{Wu2013,Gong2016} reported that \rev{the largest possible nuclear KER} is around 3 eV for two protons \rev{with} 400 nm laser pulses.
These observables at different experimental conditions show \rev{that the largest possible} nuclear KERs are around $2\sim4$ eV, which are close to our computations but far from the computations in the 2D simulations~\cite{Yue2013,Madsen2012}.
This seems to contradict to energy conservation for classical particles that the Coulomb explosion most probably starts at internuclear distance $R=2r_1=2$ atomic units, which gives nuclear KER 0.5 atomic units.
We now consider the wave packet dispersion in different coordinates.
For 1D simulation on protons with a symmetric Gaussian wave packet, the center of the wave packet moves with the velocity of the classical particle, thus \rev{the largest possible} nuclear KER is 0.5 atomic units, \rev{which is close to results in the existing 2D simulations~\cite{Yue2013,Madsen2012}}.
In 3D simulation on protons, the wave packet keeps expanding perpendicular to the polarization direction and is not symmetric \rev{on $r_1$ axis}, see Fig.~\ref{fig:wavefunctionEvolution}.
A numerical study with a simple unphysical toy model in Appendix~\ref{sec:wfExpansion} shows for the Coulomb explosion in spherical \rev{coordinates}, the most probable nuclear KER can be shifted to 1/4-1/3 of the total kinetic energy and the right half curve of the integrated JES is less steep, see Fig.~\ref{fig:wavefunctionEvolution} (a).
The longer tail in higher nuclear KER for integrated JES of $\hydroplus$ is also observed both in \rev{experiments and our} computation, but not in 1D simulation on protons, see Fig.~\ref{fig:compareRef} (d) and Fig.~\ref{fig:integratedJES}.
\rev{
Thus, the reason why we get a much lower most probable nuclear KER is that we give a 3D simulation of the wavefunction dispersion of protons, which may not be correctly approximated with 1D treatment in 2D simulations.
}The existing 2D simulations for the dissociative ionization put corrections to the electron-proton interaction with a softening parameter to give the correct ground energy of electrons $H_2^+$~\cite{Yue2013,Madsen2012}.
However, the pure Coulomb repulsion of the two protons $1/R$ ($R$ is the internuclear distance) is included without a softening parameter.
We would like to point out that, for the 2D simulation, for consistency of the correction of Coulomb interaction of the electron, the Coulomb repulsion term of the two protons may also need a softening parameter, whose value needs further investigations.
\subsection{Angular distribution}
The projected energy distribution on the azimuth angle of the electron and the protons is calculated by integrating the 6D scattering amplitudes as
\rev{
\begin{equation}
\begin{split}
 p_{N}(\theta_1, E_1)=&\int d\vec{k}_2\int d\phi_1 |b(\vec{k}_1,\vec{k}_2,T)|^2, \\
     & \vec{k}_1 = [\phi_1, \theta_1, \sqrt{8M E_1}]^T
 \end{split}
\end{equation}
for protons, and
\begin{equation}
 \begin{split}
p_{e}(\theta_2, E_2)=&\int d\vec{k}_1\int d\phi_2 |b(\vec{k}_1,\vec{k}_2,T)|^2, \\
     & \vec{k}_2 = [\phi_2, \theta_2, \sqrt{2m E_2}]^T
\end{split}
\end{equation}
}
for electron, where $E_1$ and $E_2$ are kinetic energies for {\it an individual} proton and electron.
\par
\begin{figure} 
\centering
\includegraphics[width=0.4\textwidth,trim=0.1cm 0.1cm 0.1cm 0.1cm,clip]{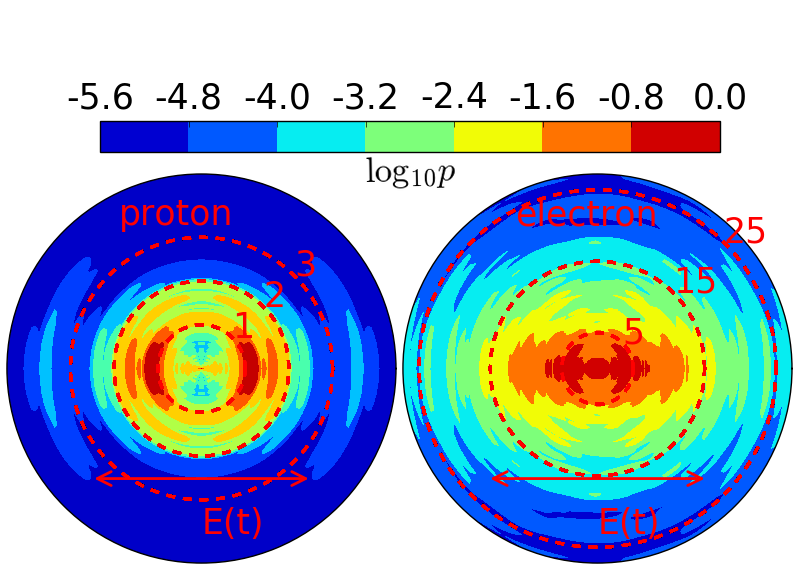}
\includegraphics[width=0.4\textwidth,trim=0.1cm 0.1cm 0.1cm 0.1cm,clip]{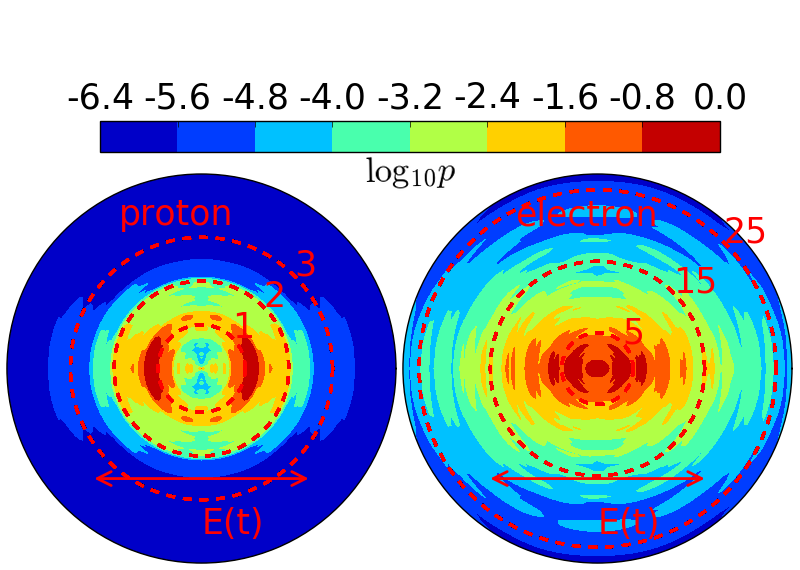}
\caption{
The log-scale probability distribution of (left column) protons by
$\log_{10}p_{N}(\theta_1,E_1)$ and (right column) protons by $\log_{10}p_{e}(\theta_2,E_2)$, $\theta_{1,2}\in[0,\pi]$.
The plot is symmetrized by $p_{N}(2\pi-\theta_1,E_1)=p_{N}(2\pi-\theta_1,E_1)$ and $p_{e}(2\pi-\theta_2,E_2)=p_{e}(2\pi-\theta_2,E_2)$.
The upper row is computed from laser pulse at intensity $\inten{8.3}{13}$ and lower row represents the $\inten{5.9}{13}$.
The values of the radial coordinates $E_{1/2}$ are represented in eV.
The polarization direction is along the horizontal axis and the direction \rev{of the} electric field is labeled at each sub-figure with an arrow and \rev{a label "$E(t)$"}.
The values are all normalized \rev{with dividing by} the maximum value.
}
\label{fig:KER}
\end{figure}

As is observed in Fig.~\ref{fig:KER}, the probability distributions of electron and protons reach the highest value in the polarization direction, which is consistent with the experimental observations for linearly polarized laser pulses~\cite{Odenweller2014,Gong2016}.
The probability of the dissociative protons is most considerable with $1\leq E_1 \leq 2$ \rev{eV}, higher than the $E_1<1$ \rev{eV} for dissociative channels reported in Ref.~\cite{Pavicic2005,Odenweller2014}, but in the range of their Coulomb explosion channel, where the laser wavelength is 800 nm.
For higher intensity $\inten{8.3}{13}$, the angular distribution of released protons and electron extends more in the polarization direction.
For distribution of protons, tiny yields around 3 eV in radial coordinates indicates the Coulomb explosion channel, close to what is observed in experiments, however, for different laser pulses~\cite{Pavicic2005}.
\section{Conclusion and discussions}
We simulate the dissociative ionization of the $\hydroplus$ ion in full dimensionality and obtain the ground energy same as the quantum chemistry methods.
Using tSurff methods, we obtained the JES where energy sharing is observed, which indicates a correlation between \rev{the electron and protons}.
The JES peaked at $E_N$ from 2 eV to 4 eV, which is different from the previous 2D simulations, but is consistent with the experimental data.
\rev{The difference indicates that in dissociative ionization of $\hydroplus$, the protons should be treated quantum mechanically in full dimensionality by simulating the 3D wavefunction evolution, where the expansion of the wave-packets perpendicular to the radial direction may need to be taken into consideration.} The projected energy distribution on angles shows that the electron and protons tend to dissociate in the direction of polarization of the laser pulse.
\par
The simulation of the single emission spectrum showing dissociation channels, is however not possible yet.
The difficulty lies mainly in constructing the internuclear-distance-dependent electronic ansatz of $H$ with a given ionic state in a single emission TDSE on $\vec{r}_1$, which might be solved by reading the energy surfaces from quantum chemistry calculations or another tRecX calculation.
\rev{This is left for future work}.
\section*{Acknowledgments}
J.Z. was supported by the DFG Priority Programme 1840, QUTIF.
We are grateful for fruitful discussions with Dr. Lun Yue from Louisiana State University, Dr. Xiaochun Gong from East China Normal University, and Prof. Dr. Armin Scrinzi from Ludwig Maximilians University.
\appendix
\renewcommand\thefigure{\thesection.\arabic{figure}}    
\section{Coordinate transformation}\label{sec:coordinates}
We use \rev{subindices  ``a'', ``b'' and ``e''} to present the two protons and the electron of an arbitrary coordinate.
The \rev{subindices ``0", ``1" and ``2"} represent the center of the two protons, the relative position of a proton to the center and the electron in our transformed coordinate, respectively.
Suppose the coordinates of the two protons and \rev{the} electron are initially represented by vectors $\vec{x}_a$,$\vec{x}_b$ and $\vec{x}_e$ of an arbitrary origin, respectively. The new coordinates $\vec{r}_1$ and $\vec{r}_2$ satisfy
\begin{equation}
\begin{split}
  \vec{r}_0=\frac{\vec{x}_a +\vec{x}_b}{2}\\
  \vec{r}_1=\frac{\vec{x}_a -\vec{x}_b }{2}\\
  \vec{r}_2=\vec{x}_e-\frac{\vec{x}_a +\vec{x}_b}{2},
\end{split}
\end{equation}
where $\vec{r}_0$ is the coordinate of the center of the two protons.
The \rev{Laplacians} of the two protons $\triangledown_a^2$,$\triangledown_b^2$ and the electron $\triangledown_e^2$ are
\begin{equation}
 \begin{split}
  \triangledown_a^2=\frac{\triangledown_0^2}{4}+\frac{\triangledown_1^2}{4}+\frac{\triangledown_2^2}{4}+\frac{\vec{\triangledown}_0\cdot\vec{\triangledown}_1}{2}
  -\frac{\vec{\triangledown}_1\cdot\vec{\triangledown}_2}{2}-\frac{\vec{\triangledown}_2\cdot\vec{\triangledown}_0}{2}\\
  \triangledown_b^2=\frac{\triangledown_0^2}{4}+\frac{\triangledown_1^2}{4}+\frac{\triangledown_2^2}{4}-\frac{\vec{\triangledown}_0\cdot\vec{\triangledown}_1}{2}
  +\frac{\vec{\triangledown}_1\cdot\vec{\triangledown}_2}{2}-\frac{\vec{\triangledown}_2\cdot\vec{\triangledown}_0}{2}\\
  \triangledown_e^2=\triangledown_2^2.
 \end{split}
\end{equation}
Thus the kinetic energy of the system can be represented by
\begin{equation}
\begin{split}
  -&\frac{1}{2}(\frac{\triangledown_a^2}{M}+\frac{\triangledown_b^2}{M}+\frac{\triangledown_e^2}{1})\\
  =&-\frac{\triangledown_0^2}{4M}-\frac{\triangledown_1^2}{4M}-\frac{\triangledown_2^2}{4M}+\frac{\vec{\triangledown}_2\cdot\vec{\triangledown}_0}{M}+\frac{\triangledown_2^2}{2}\\
  \approx &-\frac{\triangledown_1^2}{4M}-\frac{\triangledown_2^2}{2m},
\end{split}
\end{equation}
where $m=\frac{2M}{1+2M}$, and ``$\approx$" means the motion of the $\vec{r}_0$ is neglected.
The interaction of the two protons with the laser pulse can be written as
\begin{equation}
 \frac{\im}{M}\vec{A}\cdot(\vec{ \triangledown}_a+\vec{ \triangledown}_b)=\im\vec{A}\cdot\frac{1}{M}(\vec{ \triangledown}_0-\vec{ \triangledown}_2)\approx-\frac{\im}{M}\vec{A}\cdot\vec{\triangledown}_2,
\end{equation}
with which the total interaction with the laser field can be written as
\begin{equation}
 -\im\vec{A}\cdot(\vec{\triangledown}_2+\frac{1}{M}\vec{\triangledown}_2)=-\im\beta \vec{A}\cdot \vec{\triangledown}_2,
\end{equation}
where $\beta=\frac{M+1}{M}$.
\section{Convergence study}\label{sec:convergence}
\begin{figure}
\centering
\includegraphics[width=0.4\textwidth,trim=0.1cm 0.1cm 0.1cm 0.1cm,clip]{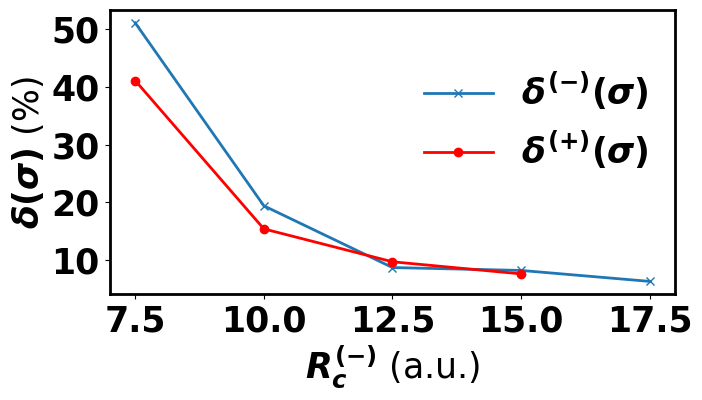}
\caption{
The (blue crosses) error $\delta^{(+)}(\sigma)$ for $R_c^{(+)}$ by Eq.~\ref{eq:convergence} with fixed $R_c^{(-)}=12.5$ atomic units and (red dots) error $\delta^{(-)}(\sigma)$ for computations for $R_c^{(-)}$ with fixed $R_c^{(+)}=12.5$ atomic units. A 400 nm laser pulse at $\inten{8.8}{13}$ is applied. For details of the laser pulse, refer to Sec.~\ref{sec:pulses}. The parameters for angular momenta are $L_{\max}=7$, $M_{\max}=5$.
}
\label{fig:convergence}
\end{figure}

The errors are computed by the difference of JES from two subsequent calculations $\sigma(E_N,E_e)$ and $\sigma'(E_N,E_e)$ with respect to $R_c^{(+)}$ and $R_c^{(-)}$
\begin{equation}\label{eq:convergence}
 \delta(\sigma)= \max_{E_N,E_e}2\frac{|\sigma(E_N,E_e)-\sigma'(E_N,E_e)|}{|\sigma(E_N,E_e)+\sigma'(E_N,E_e)|},
\end{equation}
as used in previously in Fig.~\ref{fig:H2PlusJES} (c).
As depicted in Fig.~\ref{fig:convergence}, the JES is converged at $R_c^{(+)}=R_c^{(-)}=12.5$ atomic units with error below 10\%.
\section{Coulomb explosion with dispersion of wave packets}\label{sec:wfExpansion}
\rev{Here, we illustrate} the dispersion of the wavefunction with $r$ in spherical \rev{coordinates} by a simple but unphysical model: the Coulomb explosion of a hydrogen atom.
It starts with the ground eigenstate of a hydrogen atom with electronic wavefunction $\psi_0(\vec{r})=\psi_0(r)=\frac{1}{\sqrt{\pi} }\exp(-r)$.
The initial kinetic energy of the electron is 0.5 atomic units and the potential energy is -1 atomic units.
Then, the charge of the nucleus suddenly changes from +1 to -1 and the kinetic energy remains the same, but the Coulomb potential reverses its sign.
Thus the system explodes and the electron finally becomes a free particle.
\rev{In what follows, we will discuss the largest possible kinetic energy of the free electron.}
\par
\begin{figure}
\centering
\includegraphics[width=0.4\textwidth,trim=0.1cm 0.1cm 0.1cm 0.1cm,clip]{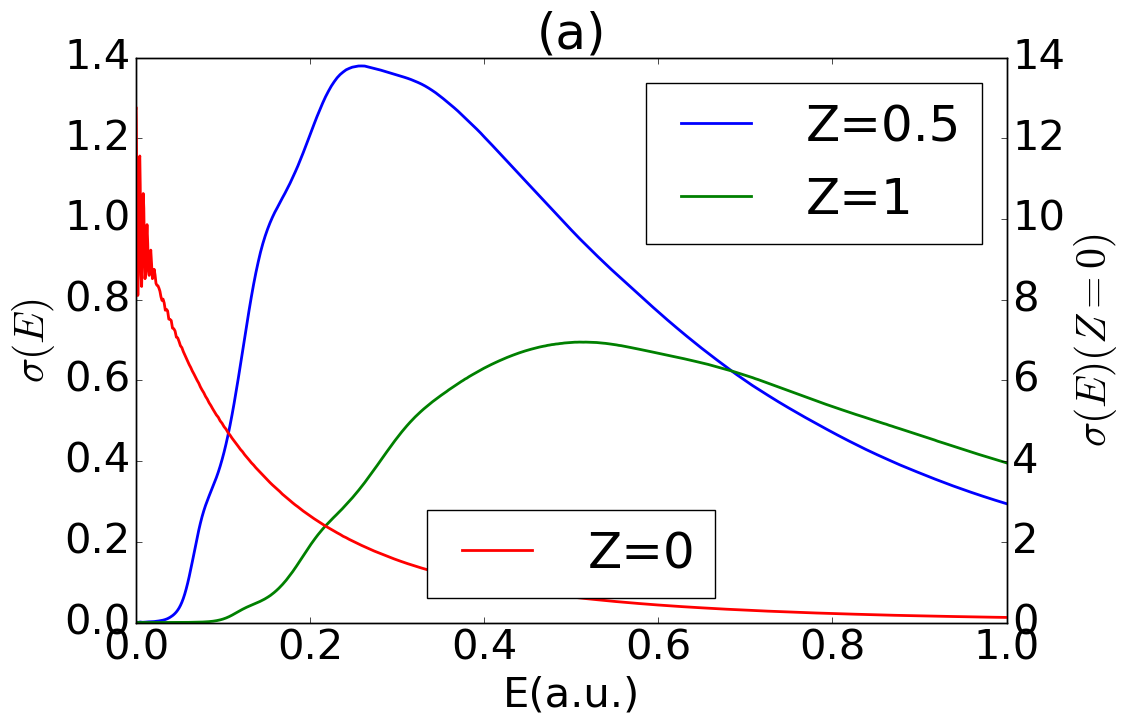}
\includegraphics[width=0.4\textwidth,trim=0.1cm 0.1cm 0.1cm 0.1cm,clip]{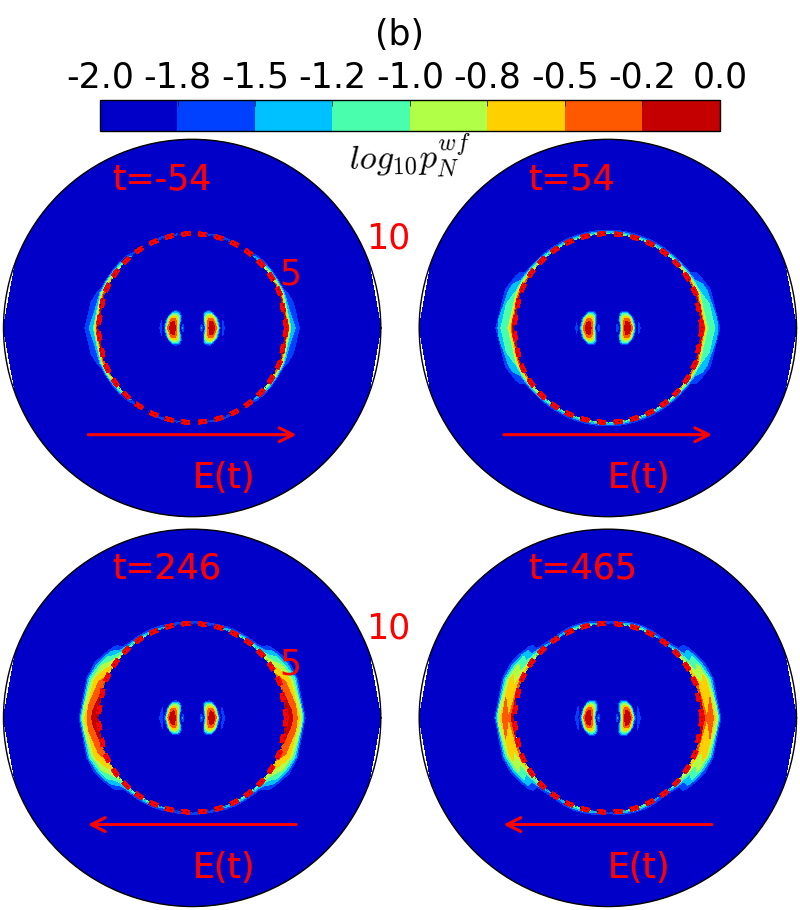}
\caption{
(a) The energy spectrum $\sigma(E)=\int d\theta \int d\phi k |b(\vec{k},T)|^2,\,E=\frac{k^2}{2}$, where $b(\vec{k},T)$ are the single electron scattering amplitudes. The spectrum is computed \rev{by} advancing a hydrogen electronic ground state in the Coulomb potential $+\frac{Z}{r}$. Z=0 means no external Coulomb potential.
The kinetic energy \rev{is computed} by $\int_0^{\infty}\sigma(E)\sqrt{E}dE=Z+0.5$ atomic units, consistent with energy conservation.
(b) The log-scale probability distribution of protons \rev{is computed} by
$\log_{10}p^{wf}_{N}(\phi_1,\theta_1, r_1,0,5, t), r_1\in [0,5]$ of the inner shell and 
$\log_{10}p^{wf}_{N}(\phi_1,\theta_1, r_1, 5, 10, t), r_1\in [5,10]$ of the outer shell.
\rev{$|E_z(t)|$ is considerable and reaches a local maximum when t=-54, 54 and 246 atomic units, which are also depicted in the figures. When t=465 atomic units, $|E_z(t)|$ is very small.
The whole space is split into two regions by the red, dashed circle, and the signals of each region are normalized independently.} The polarization direction is along the horizontal axis and the direction \rev{of the} electric field is labeled at each sub-figure with an arrow \rev{and a label ``$E(t)$"}.
The absolute value of the outer shell is several orders smaller than the inner shell.
}
\label{fig:wavefunctionEvolution}
\end{figure}

 In classical mechanics, the \rev{largest possible kinetic energy} is 1.5 atomic units because of energy conservation.
However, our quantum simulation by tSurff gives a \rev{value} around 0.5 atomic units, see Fig.~\ref{fig:wavefunctionEvolution} (a).
We can also see that the distribution of the spectrum has long tails to \rev{the} high energy region.
The integration of $\sigma(E)$ gives the total energy 1.5 atomic units, which is consistent with energy conservation, see Fig.~\ref{fig:wavefunctionEvolution} (a).

\par
The numerical simulations show the nuclear wave packets also expand in space during the time propagation, see Fig.~\ref{fig:wavefunctionEvolution} (b), where the probability \rev{distribution} of the protons \rev{over} time is computed by
\begin{equation}
\begin{split}
 p^{wf}_{N}(\phi_1,\theta_1, r_1, R_0,R_1, t)=&\int d\phi_2 \int \sin\theta_2 d\theta_2\\
 &\int_{R_0}^{R_1} r_2^2 dr_2|\psi(\vec{r}_1,\vec{r}_2,t)|^2.
 \end{split}
\end{equation}
We split the radial coordinates into inner region $r_1,r_2\in [R_0,R_1]=[0,5]$ and outer region $r_1,r_2 \in [R_0,R_1]=[5,10]$, and the yields of both regions are normalized \rev{with diving by} the maximum probability of the region.
Thus we attribute the difference of most possible nuclear KER in JES to the 3D wavefunction dispersion and expansion, which \rev{were} not included in previous simulations.
\begin{figure}
\centering
\includegraphics[width=0.4\textwidth,trim=0.1cm 0.1cm 0.1cm 0.1cm,clip]{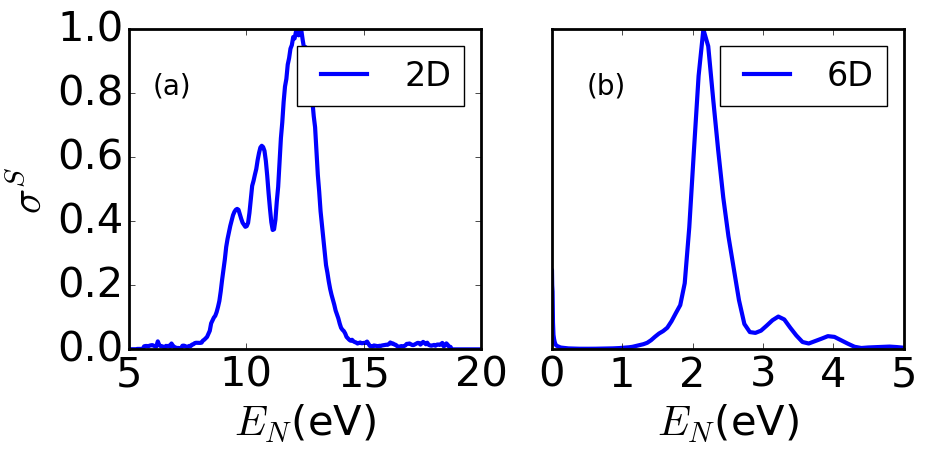}
\caption{The normalized single JES is created from (a) 2D and (b) 6D $\sigma^S(E_N)=\int \sigma(2E_N,E_{e}) \sqrt{E_e} dE_e$, where $E_N$ depicts the kinetic energy of a proton. A laser pulse used in Ref.~\cite{Yue2013} is applied here. The 2D data is taken from Ref.~\cite{Yue2013}.
}
\label{fig:integratedJES}
\end{figure}
\bibliography{h2plus.bib}

\begin{thebibliography}{39}%
\makeatletter
\providecommand \@ifxundefined [1]{%
 \@ifx{#1\undefined}
}%
\providecommand \@ifnum [1]{%
 \ifnum #1\expandafter \@firstoftwo
 \else \expandafter \@secondoftwo
 \fi
}%
\providecommand \@ifx [1]{%
 \ifx #1\expandafter \@firstoftwo
 \else \expandafter \@secondoftwo
 \fi
}%
\providecommand \natexlab [1]{#1}%
\providecommand \enquote  [1]{``#1''}%
\providecommand \bibnamefont  [1]{#1}%
\providecommand \bibfnamefont [1]{#1}%
\providecommand \citenamefont [1]{#1}%
\providecommand \href@noop [0]{\@secondoftwo}%
\providecommand \href [0]{\begingroup \@sanitize@url \@href}%
\providecommand \@href[1]{\@@startlink{#1}\@@href}%
\providecommand \@@href[1]{\endgroup#1\@@endlink}%
\providecommand \@sanitize@url [0]{\catcode `\\12\catcode `\$12\catcode
  `\&12\catcode `\#12\catcode `\^12\catcode `\_12\catcode `\%12\relax}%
\providecommand \@@startlink[1]{}%
\providecommand \@@endlink[0]{}%
\providecommand \url  [0]{\begingroup\@sanitize@url \@url }%
\providecommand \@url [1]{\endgroup\@href {#1}{\urlprefix }}%
\providecommand \urlprefix  [0]{URL }%
\providecommand \Eprint [0]{\href }%
\providecommand \doibase [0]{http://dx.doi.org/}%
\providecommand \selectlanguage [0]{\@gobble}%
\providecommand \bibinfo  [0]{\@secondoftwo}%
\providecommand \bibfield  [0]{\@secondoftwo}%
\providecommand \translation [1]{[#1]}%
\providecommand \BibitemOpen [0]{}%
\providecommand \bibitemStop [0]{}%
\providecommand \bibitemNoStop [0]{.\EOS\space}%
\providecommand \EOS [0]{\spacefactor3000\relax}%
\providecommand \BibitemShut  [1]{\csname bibitem#1\endcsname}%
\let\auto@bib@innerbib\@empty
\bibitem [{\citenamefont {Bucksbaum}\ \emph {et~al.}(1990)\citenamefont
  {Bucksbaum}, \citenamefont {Zavriyev}, \citenamefont {Muller},\ and\
  \citenamefont {Schumacher}}]{Bucksbaum1990}%
  \BibitemOpen
  \bibfield  {author} {\bibinfo {author} {\bibfnamefont {P.~H.}\ \bibnamefont
  {Bucksbaum}}, \bibinfo {author} {\bibfnamefont {A.}~\bibnamefont {Zavriyev}},
  \bibinfo {author} {\bibfnamefont {H.~G.}\ \bibnamefont {Muller}}, \ and\
  \bibinfo {author} {\bibfnamefont {D.~W.}\ \bibnamefont {Schumacher}},\ }\href
  {\doibase 10.1103/PhysRevLett.64.1883} {\bibfield  {journal} {\bibinfo
  {journal} {Phys. Rev. Lett.}\ }\textbf {\bibinfo {volume} {64}},\ \bibinfo
  {pages} {1883} (\bibinfo {year} {1990})}\BibitemShut {NoStop}%
\bibitem [{\citenamefont {Zuo}\ and\ \citenamefont {Bandrauk}(1995)}]{Zuo1995}%
  \BibitemOpen
  \bibfield  {author} {\bibinfo {author} {\bibfnamefont {T.}~\bibnamefont
  {Zuo}}\ and\ \bibinfo {author} {\bibfnamefont {A.~D.}\ \bibnamefont
  {Bandrauk}},\ }\href {\doibase 10.1103/PhysRevA.52.R2511} {\bibfield
  {journal} {\bibinfo  {journal} {Phys. Rev. A}\ }\textbf {\bibinfo {volume}
  {52}},\ \bibinfo {pages} {R2511} (\bibinfo {year} {1995})}\BibitemShut
  {NoStop}%
\bibitem [{\citenamefont {Yao}\ and\ \citenamefont {Chu}(1993)}]{Yao1993}%
  \BibitemOpen
  \bibfield  {author} {\bibinfo {author} {\bibfnamefont {G.}~\bibnamefont
  {Yao}}\ and\ \bibinfo {author} {\bibfnamefont {S.-I.}\ \bibnamefont {Chu}},\
  }\href {\doibase 10.1103/PhysRevA.48.485} {\bibfield  {journal} {\bibinfo
  {journal} {Phys. Rev. A}\ }\textbf {\bibinfo {volume} {48}},\ \bibinfo
  {pages} {485} (\bibinfo {year} {1993})}\BibitemShut {NoStop}%
\bibitem [{\citenamefont {Giusti-Suzor}\ \emph {et~al.}(1995)\citenamefont
  {Giusti-Suzor}, \citenamefont {Mies}, \citenamefont {DiMauro}, \citenamefont
  {Charron},\ and\ \citenamefont {Yang}}]{Giusti-Suzor1995}%
  \BibitemOpen
  \bibfield  {author} {\bibinfo {author} {\bibfnamefont {A.}~\bibnamefont
  {Giusti-Suzor}}, \bibinfo {author} {\bibfnamefont {F.~H.}\ \bibnamefont
  {Mies}}, \bibinfo {author} {\bibfnamefont {L.~F.}\ \bibnamefont {DiMauro}},
  \bibinfo {author} {\bibfnamefont {E.}~\bibnamefont {Charron}}, \ and\
  \bibinfo {author} {\bibfnamefont {B.}~\bibnamefont {Yang}},\ }\href {\doibase
  10.1088/0953-4075/28/3/006} {\bibfield  {journal} {\bibinfo  {journal} {J.
  Phys. B At. Mol. Opt. Phys.}\ }\textbf {\bibinfo {volume} {28}},\ \bibinfo
  {pages} {309} (\bibinfo {year} {1995})}\BibitemShut {NoStop}%
\bibitem [{\citenamefont {Jolicard}\ and\ \citenamefont
  {Atabek}(1992)}]{Jolicard1992}%
  \BibitemOpen
  \bibfield  {author} {\bibinfo {author} {\bibfnamefont {G.}~\bibnamefont
  {Jolicard}}\ and\ \bibinfo {author} {\bibfnamefont {O.}~\bibnamefont
  {Atabek}},\ }\href {\doibase 10.1103/PhysRevA.46.5845} {\bibfield  {journal}
  {\bibinfo  {journal} {Phys. Rev. A}\ }\textbf {\bibinfo {volume} {46}},\
  \bibinfo {pages} {5845} (\bibinfo {year} {1992})}\BibitemShut {NoStop}%
\bibitem [{\citenamefont {Zuo}\ \emph {et~al.}(1993)\citenamefont {Zuo},
  \citenamefont {Chelkowski},\ and\ \citenamefont {Bandrauk}}]{Zuo1993}%
  \BibitemOpen
  \bibfield  {author} {\bibinfo {author} {\bibfnamefont {T.}~\bibnamefont
  {Zuo}}, \bibinfo {author} {\bibfnamefont {S.}~\bibnamefont {Chelkowski}}, \
  and\ \bibinfo {author} {\bibfnamefont {A.~D.}\ \bibnamefont {Bandrauk}},\
  }\href {\doibase 10.1103/PhysRevA.48.3837} {\bibfield  {journal} {\bibinfo
  {journal} {Phys. Rev. A}\ }\textbf {\bibinfo {volume} {48}},\ \bibinfo
  {pages} {3837} (\bibinfo {year} {1993})}\BibitemShut {NoStop}%
\bibitem [{\citenamefont {Esry}\ \emph {et~al.}(2006)\citenamefont {Esry},
  \citenamefont {Sayler}, \citenamefont {Wang}, \citenamefont {Carnes},\ and\
  \citenamefont {Ben-Itzhak}}]{Esry2006}%
  \BibitemOpen
  \bibfield  {author} {\bibinfo {author} {\bibfnamefont {B.~D.}\ \bibnamefont
  {Esry}}, \bibinfo {author} {\bibfnamefont {A.~M.}\ \bibnamefont {Sayler}},
  \bibinfo {author} {\bibfnamefont {P.~Q.}\ \bibnamefont {Wang}}, \bibinfo
  {author} {\bibfnamefont {K.~D.}\ \bibnamefont {Carnes}}, \ and\ \bibinfo
  {author} {\bibfnamefont {I.}~\bibnamefont {Ben-Itzhak}},\ }\href {\doibase
  10.1103/PhysRevLett.97.013003} {\bibfield  {journal} {\bibinfo  {journal}
  {Phys. Rev. Lett.}\ }\textbf {\bibinfo {volume} {97}},\ \bibinfo {pages}
  {013003} (\bibinfo {year} {2006})}\BibitemShut {NoStop}%
\bibitem [{\citenamefont {Posthumus}(2004)}]{Posthumus2004}%
  \BibitemOpen
  \bibfield  {author} {\bibinfo {author} {\bibfnamefont {J.~H.}\ \bibnamefont
  {Posthumus}},\ }\href {\doibase 10.1088/0034-4885/67/5/R01} {\bibfield
  {journal} {\bibinfo  {journal} {Reports Prog. Phys.}\ }\textbf {\bibinfo
  {volume} {67}},\ \bibinfo {pages} {623} (\bibinfo {year} {2004})}\BibitemShut
  {NoStop}%
\bibitem [{\citenamefont {Giusti-Suzor}\ \emph {et~al.}(1990)\citenamefont
  {Giusti-Suzor}, \citenamefont {He}, \citenamefont {Atabek},\ and\
  \citenamefont {Mies}}]{Giusti-Suzor1990}%
  \BibitemOpen
  \bibfield  {author} {\bibinfo {author} {\bibfnamefont {A.}~\bibnamefont
  {Giusti-Suzor}}, \bibinfo {author} {\bibfnamefont {X.}~\bibnamefont {He}},
  \bibinfo {author} {\bibfnamefont {O.}~\bibnamefont {Atabek}}, \ and\ \bibinfo
  {author} {\bibfnamefont {F.~H.}\ \bibnamefont {Mies}},\ }\href {\doibase
  10.1103/PhysRevLett.64.515} {\bibfield  {journal} {\bibinfo  {journal} {Phys.
  Rev. Lett.}\ }\textbf {\bibinfo {volume} {64}},\ \bibinfo {pages} {515}
  (\bibinfo {year} {1990})}\BibitemShut {NoStop}%
\bibitem [{\citenamefont {Odenweller}\ \emph {et~al.}(2011)\citenamefont
  {Odenweller}, \citenamefont {Takemoto}, \citenamefont {Vredenborg},
  \citenamefont {Cole}, \citenamefont {Pahl}, \citenamefont {Titze},
  \citenamefont {Schmidt}, \citenamefont {Jahnke}, \citenamefont
  {D{\"{o}}rner},\ and\ \citenamefont {Becker}}]{Odenweller2011}%
  \BibitemOpen
  \bibfield  {author} {\bibinfo {author} {\bibfnamefont {M.}~\bibnamefont
  {Odenweller}}, \bibinfo {author} {\bibfnamefont {N.}~\bibnamefont
  {Takemoto}}, \bibinfo {author} {\bibfnamefont {A.}~\bibnamefont
  {Vredenborg}}, \bibinfo {author} {\bibfnamefont {K.}~\bibnamefont {Cole}},
  \bibinfo {author} {\bibfnamefont {K.}~\bibnamefont {Pahl}}, \bibinfo {author}
  {\bibfnamefont {J.}~\bibnamefont {Titze}}, \bibinfo {author} {\bibfnamefont
  {L.~P.~H.}\ \bibnamefont {Schmidt}}, \bibinfo {author} {\bibfnamefont
  {T.}~\bibnamefont {Jahnke}}, \bibinfo {author} {\bibfnamefont
  {R.}~\bibnamefont {D{\"{o}}rner}}, \ and\ \bibinfo {author} {\bibfnamefont
  {A.}~\bibnamefont {Becker}},\ }\href {\doibase
  10.1103/PhysRevLett.107.143004} {\bibfield  {journal} {\bibinfo  {journal}
  {Phys. Rev. Lett.}\ }\textbf {\bibinfo {volume} {107}},\ \bibinfo {pages}
  {143004} (\bibinfo {year} {2011})}\BibitemShut {NoStop}%
\bibitem [{\citenamefont {Odenweller}\ \emph {et~al.}(2014)\citenamefont
  {Odenweller}, \citenamefont {Lower}, \citenamefont {Pahl}, \citenamefont
  {Sch{\"{u}}tt}, \citenamefont {Wu}, \citenamefont {Cole}, \citenamefont
  {Vredenborg}, \citenamefont {Schmidt}, \citenamefont {Neumann}, \citenamefont
  {Titze}, \citenamefont {Jahnke}, \citenamefont {Meckel}, \citenamefont
  {Kunitski}, \citenamefont {Havermeier}, \citenamefont {Voss}, \citenamefont
  {Sch{\"{o}}ffler}, \citenamefont {Sann}, \citenamefont {Voigtsberger},
  \citenamefont {Schmidt-B{\"{o}}cking},\ and\ \citenamefont
  {D{\"{o}}rner}}]{Odenweller2014}%
  \BibitemOpen
  \bibfield  {author} {\bibinfo {author} {\bibfnamefont {M.}~\bibnamefont
  {Odenweller}}, \bibinfo {author} {\bibfnamefont {J.}~\bibnamefont {Lower}},
  \bibinfo {author} {\bibfnamefont {K.}~\bibnamefont {Pahl}}, \bibinfo {author}
  {\bibfnamefont {M.}~\bibnamefont {Sch{\"{u}}tt}}, \bibinfo {author}
  {\bibfnamefont {J.}~\bibnamefont {Wu}}, \bibinfo {author} {\bibfnamefont
  {K.}~\bibnamefont {Cole}}, \bibinfo {author} {\bibfnamefont {A.}~\bibnamefont
  {Vredenborg}}, \bibinfo {author} {\bibfnamefont {L.~P.}\ \bibnamefont
  {Schmidt}}, \bibinfo {author} {\bibfnamefont {N.}~\bibnamefont {Neumann}},
  \bibinfo {author} {\bibfnamefont {J.}~\bibnamefont {Titze}}, \bibinfo
  {author} {\bibfnamefont {T.}~\bibnamefont {Jahnke}}, \bibinfo {author}
  {\bibfnamefont {M.}~\bibnamefont {Meckel}}, \bibinfo {author} {\bibfnamefont
  {M.}~\bibnamefont {Kunitski}}, \bibinfo {author} {\bibfnamefont
  {T.}~\bibnamefont {Havermeier}}, \bibinfo {author} {\bibfnamefont
  {S.}~\bibnamefont {Voss}}, \bibinfo {author} {\bibfnamefont {M.}~\bibnamefont
  {Sch{\"{o}}ffler}}, \bibinfo {author} {\bibfnamefont {H.}~\bibnamefont
  {Sann}}, \bibinfo {author} {\bibfnamefont {J.}~\bibnamefont {Voigtsberger}},
  \bibinfo {author} {\bibfnamefont {H.}~\bibnamefont {Schmidt-B{\"{o}}cking}},
  \ and\ \bibinfo {author} {\bibfnamefont {R.}~\bibnamefont {D{\"{o}}rner}},\
  }\href {\doibase 10.1103/PhysRevA.89.013424} {\bibfield  {journal} {\bibinfo
  {journal} {Phys. Rev. A - At. Mol. Opt. Phys.}\ }\textbf {\bibinfo {volume}
  {89}},\ \bibinfo {pages} {013424} (\bibinfo {year} {2014})}\BibitemShut
  {NoStop}%
\bibitem [{\citenamefont {Wu}\ \emph {et~al.}(2013)\citenamefont {Wu},
  \citenamefont {Kunitski}, \citenamefont {Pitzer}, \citenamefont {Trinter},
  \citenamefont {Schmidt}, \citenamefont {Jahnke}, \citenamefont
  {Magrakvelidze}, \citenamefont {Madsen}, \citenamefont {Madsen},
  \citenamefont {Thumm},\ and\ \citenamefont {D{\"{o}}rner}}]{Wu2013}%
  \BibitemOpen
  \bibfield  {author} {\bibinfo {author} {\bibfnamefont {J.}~\bibnamefont
  {Wu}}, \bibinfo {author} {\bibfnamefont {M.}~\bibnamefont {Kunitski}},
  \bibinfo {author} {\bibfnamefont {M.}~\bibnamefont {Pitzer}}, \bibinfo
  {author} {\bibfnamefont {F.}~\bibnamefont {Trinter}}, \bibinfo {author}
  {\bibfnamefont {L.~P.~H.}\ \bibnamefont {Schmidt}}, \bibinfo {author}
  {\bibfnamefont {T.}~\bibnamefont {Jahnke}}, \bibinfo {author} {\bibfnamefont
  {M.}~\bibnamefont {Magrakvelidze}}, \bibinfo {author} {\bibfnamefont {C.~B.}\
  \bibnamefont {Madsen}}, \bibinfo {author} {\bibfnamefont {L.~B.}\
  \bibnamefont {Madsen}}, \bibinfo {author} {\bibfnamefont {U.}~\bibnamefont
  {Thumm}}, \ and\ \bibinfo {author} {\bibfnamefont {R.}~\bibnamefont
  {D{\"{o}}rner}},\ }\href {\doibase 10.1103/PhysRevLett.111.023002} {\bibfield
   {journal} {\bibinfo  {journal} {Phys. Rev. Lett.}\ }\textbf {\bibinfo
  {volume} {111}},\ \bibinfo {pages} {023002} (\bibinfo {year}
  {2013})}\BibitemShut {NoStop}%
\bibitem [{\citenamefont {Gong}\ \emph {et~al.}(2016)\citenamefont {Gong},
  \citenamefont {He}, \citenamefont {Song}, \citenamefont {Ji}, \citenamefont
  {Lin}, \citenamefont {Zhang}, \citenamefont {Lu}, \citenamefont {Pan},
  \citenamefont {Ding}, \citenamefont {Zeng}, \citenamefont {He},\ and\
  \citenamefont {Wu}}]{Gong2016}%
  \BibitemOpen
  \bibfield  {author} {\bibinfo {author} {\bibfnamefont {X.}~\bibnamefont
  {Gong}}, \bibinfo {author} {\bibfnamefont {P.}~\bibnamefont {He}}, \bibinfo
  {author} {\bibfnamefont {Q.}~\bibnamefont {Song}}, \bibinfo {author}
  {\bibfnamefont {Q.}~\bibnamefont {Ji}}, \bibinfo {author} {\bibfnamefont
  {K.}~\bibnamefont {Lin}}, \bibinfo {author} {\bibfnamefont {W.}~\bibnamefont
  {Zhang}}, \bibinfo {author} {\bibfnamefont {P.}~\bibnamefont {Lu}}, \bibinfo
  {author} {\bibfnamefont {H.}~\bibnamefont {Pan}}, \bibinfo {author}
  {\bibfnamefont {J.}~\bibnamefont {Ding}}, \bibinfo {author} {\bibfnamefont
  {H.}~\bibnamefont {Zeng}}, \bibinfo {author} {\bibfnamefont {F.}~\bibnamefont
  {He}}, \ and\ \bibinfo {author} {\bibfnamefont {J.}~\bibnamefont {Wu}},\
  }\href {\doibase 10.1364/optica.3.000643} {\bibfield  {journal} {\bibinfo
  {journal} {Optica}\ }\textbf {\bibinfo {volume} {3}},\ \bibinfo {pages} {643}
  (\bibinfo {year} {2016})}\BibitemShut {NoStop}%
\bibitem [{\citenamefont {Madsen}\ \emph {et~al.}(2012)\citenamefont {Madsen},
  \citenamefont {Anis}, \citenamefont {Madsen},\ and\ \citenamefont
  {Esry}}]{Madsen2012}%
  \BibitemOpen
  \bibfield  {author} {\bibinfo {author} {\bibfnamefont {C.~B.}\ \bibnamefont
  {Madsen}}, \bibinfo {author} {\bibfnamefont {F.}~\bibnamefont {Anis}},
  \bibinfo {author} {\bibfnamefont {L.~B.}\ \bibnamefont {Madsen}}, \ and\
  \bibinfo {author} {\bibfnamefont {B.~D.}\ \bibnamefont {Esry}},\ }\href
  {\doibase 10.1103/PhysRevLett.109.163003} {\bibfield  {journal} {\bibinfo
  {journal} {Phys. Rev. Lett.}\ }\textbf {\bibinfo {volume} {109}},\ \bibinfo
  {pages} {163003} (\bibinfo {year} {2012})}\BibitemShut {NoStop}%
\bibitem [{\citenamefont {Scrinzi}(2012)}]{Scrinzi2012}%
  \BibitemOpen
  \bibfield  {author} {\bibinfo {author} {\bibfnamefont {A.}~\bibnamefont
  {Scrinzi}},\ }\href {\doibase 10.1088/1367-2630/14/8/085008} {\bibfield
  {journal} {\bibinfo  {journal} {New J. Phys.}\ }\textbf {\bibinfo {volume}
  {14}},\ \bibinfo {pages} {085008} (\bibinfo {year} {2012})}\BibitemShut
  {NoStop}%
\bibitem [{\citenamefont {Zielinski}\ \emph {et~al.}(2016)\citenamefont
  {Zielinski}, \citenamefont {Majety},\ and\ \citenamefont
  {Scrinzi}}]{Zielinski2016}%
  \BibitemOpen
  \bibfield  {author} {\bibinfo {author} {\bibfnamefont {A.}~\bibnamefont
  {Zielinski}}, \bibinfo {author} {\bibfnamefont {V.~P.}\ \bibnamefont
  {Majety}}, \ and\ \bibinfo {author} {\bibfnamefont {A.}~\bibnamefont
  {Scrinzi}},\ }\href {\doibase 10.1103/PhysRevA.93.023406} {\bibfield
  {journal} {\bibinfo  {journal} {Phys. Rev. A}\ }\textbf {\bibinfo {volume}
  {93}},\ \bibinfo {pages} {023406} (\bibinfo {year} {2016})}\BibitemShut
  {NoStop}%
\bibitem [{\citenamefont {Tao}\ and\ \citenamefont {Scrinzi}(2012)}]{Tao2012}%
  \BibitemOpen
  \bibfield  {author} {\bibinfo {author} {\bibfnamefont {L.}~\bibnamefont
  {Tao}}\ and\ \bibinfo {author} {\bibfnamefont {A.}~\bibnamefont {Scrinzi}},\
  }\href {\doibase 10.1088/1367-2630/14/1/013021} {\bibfield  {journal}
  {\bibinfo  {journal} {New J. Phys.}\ }\textbf {\bibinfo {volume} {14}},\
  \bibinfo {pages} {013021} (\bibinfo {year} {2012})}\BibitemShut {NoStop}%
\bibitem [{\citenamefont {Zhu}\ and\ \citenamefont {Scrinzi}(2020)}]{Zhu2020}%
  \BibitemOpen
  \bibfield  {author} {\bibinfo {author} {\bibfnamefont {J.}~\bibnamefont
  {Zhu}}\ and\ \bibinfo {author} {\bibfnamefont {A.}~\bibnamefont {Scrinzi}},\
  }\href {\doibase 10.1103/PhysRevA.101.063407} {\bibfield  {journal} {\bibinfo
   {journal} {Phys. Rev. A}\ }\textbf {\bibinfo {volume} {101}},\ \bibinfo
  {pages} {063407} (\bibinfo {year} {2020})}\BibitemShut {NoStop}%
\bibitem [{\citenamefont {Yue}\ and\ \citenamefont {Madsen}(2013)}]{Yue2013}%
  \BibitemOpen
  \bibfield  {author} {\bibinfo {author} {\bibfnamefont {L.}~\bibnamefont
  {Yue}}\ and\ \bibinfo {author} {\bibfnamefont {L.~B.}\ \bibnamefont
  {Madsen}},\ }\href {\doibase 10.1103/PhysRevA.88.063420} {\bibfield
  {journal} {\bibinfo  {journal} {Phys. Rev. A}\ }\textbf {\bibinfo {volume}
  {88}},\ \bibinfo {pages} {063420} (\bibinfo {year} {2013})}\BibitemShut
  {NoStop}%
\bibitem [{\citenamefont {Yue}\ and\ \citenamefont {Madsen}(2014)}]{Yue2014}%
  \BibitemOpen
  \bibfield  {author} {\bibinfo {author} {\bibfnamefont {L.}~\bibnamefont
  {Yue}}\ and\ \bibinfo {author} {\bibfnamefont {L.~B.}\ \bibnamefont
  {Madsen}},\ }\href {\doibase 10.1103/PhysRevA.90.063408} {\bibfield
  {journal} {\bibinfo  {journal} {Phys. Rev. A}\ }\textbf {\bibinfo {volume}
  {90}},\ \bibinfo {pages} {063408} (\bibinfo {year} {2014})}\BibitemShut
  {NoStop}%
\bibitem [{\citenamefont {Steeg}\ \emph {et~al.}(2003)\citenamefont {Steeg},
  \citenamefont {Bartschat},\ and\ \citenamefont {Bray}}]{Steeg2003}%
  \BibitemOpen
  \bibfield  {author} {\bibinfo {author} {\bibfnamefont {G.~L.~V.}\
  \bibnamefont {Steeg}}, \bibinfo {author} {\bibfnamefont {K.}~\bibnamefont
  {Bartschat}}, \ and\ \bibinfo {author} {\bibfnamefont {I.}~\bibnamefont
  {Bray}},\ }\href {\doibase 10.1088/0953-4075/36/15/313} {\bibfield  {journal}
  {\bibinfo  {journal} {J. Phys. B At. Mol. Opt. Phys.}\ }\textbf {\bibinfo
  {volume} {36}},\ \bibinfo {pages} {3325} (\bibinfo {year}
  {2003})}\BibitemShut {NoStop}%
\bibitem [{\citenamefont {Qu}\ \emph {et~al.}(2001)\citenamefont {Qu},
  \citenamefont {Chen}, \citenamefont {Xu},\ and\ \citenamefont
  {Keitel}}]{Qu2002}%
  \BibitemOpen
  \bibfield  {author} {\bibinfo {author} {\bibfnamefont {W.}~\bibnamefont
  {Qu}}, \bibinfo {author} {\bibfnamefont {Z.}~\bibnamefont {Chen}}, \bibinfo
  {author} {\bibfnamefont {Z.}~\bibnamefont {Xu}}, \ and\ \bibinfo {author}
  {\bibfnamefont {C.~H.}\ \bibnamefont {Keitel}},\ }\href {\doibase
  10.1103/PhysRevA.65.013402} {\bibfield  {journal} {\bibinfo  {journal} {Phys.
  Rev. A}\ }\textbf {\bibinfo {volume} {65}},\ \bibinfo {pages} {013402}
  (\bibinfo {year} {2001})}\BibitemShut {NoStop}%
\bibitem [{\citenamefont {Silva}\ \emph {et~al.}(2013)\citenamefont {Silva},
  \citenamefont {Catoire}, \citenamefont {Rivi{\`{e}}re}, \citenamefont
  {Bachau},\ and\ \citenamefont {Mart{\'{i}}n}}]{Silva2013}%
  \BibitemOpen
  \bibfield  {author} {\bibinfo {author} {\bibfnamefont {R.~E.~F.}\
  \bibnamefont {Silva}}, \bibinfo {author} {\bibfnamefont {F.}~\bibnamefont
  {Catoire}}, \bibinfo {author} {\bibfnamefont {P.}~\bibnamefont
  {Rivi{\`{e}}re}}, \bibinfo {author} {\bibfnamefont {H.}~\bibnamefont
  {Bachau}}, \ and\ \bibinfo {author} {\bibfnamefont {F.}~\bibnamefont
  {Mart{\'{i}}n}},\ }\href {\doibase 10.1103/PhysRevLett.110.113001} {\bibfield
   {journal} {\bibinfo  {journal} {Phys. Rev. Lett.}\ }\textbf {\bibinfo
  {volume} {110}},\ \bibinfo {pages} {113001} (\bibinfo {year}
  {2013})}\BibitemShut {NoStop}%
\bibitem [{\citenamefont {Takemoto}\ and\ \citenamefont
  {Becker}(2010)}]{Takemoto2010}%
  \BibitemOpen
  \bibfield  {author} {\bibinfo {author} {\bibfnamefont {N.}~\bibnamefont
  {Takemoto}}\ and\ \bibinfo {author} {\bibfnamefont {A.}~\bibnamefont
  {Becker}},\ }\href {\doibase 10.1103/PhysRevLett.105.203004} {\bibfield
  {journal} {\bibinfo  {journal} {Phys. Rev. Lett.}\ }\textbf {\bibinfo
  {volume} {105}},\ \bibinfo {pages} {203004} (\bibinfo {year}
  {2010})}\BibitemShut {NoStop}%
\bibitem [{\citenamefont {Feuerstein}\ and\ \citenamefont
  {Thumm}(2003)}]{Feuerstein2003}%
  \BibitemOpen
  \bibfield  {author} {\bibinfo {author} {\bibfnamefont {B.}~\bibnamefont
  {Feuerstein}}\ and\ \bibinfo {author} {\bibfnamefont {U.}~\bibnamefont
  {Thumm}},\ }\href {\doibase 10.1103/PhysRevA.67.043405} {\bibfield  {journal}
  {\bibinfo  {journal} {Phys. Rev. A - At. Mol. Opt. Phys.}\ }\textbf {\bibinfo
  {volume} {67}},\ \bibinfo {pages} {043405} (\bibinfo {year}
  {2003})}\BibitemShut {NoStop}%
\bibitem [{\citenamefont {Kulander}\ \emph {et~al.}(1996)\citenamefont
  {Kulander}, \citenamefont {Mies},\ and\ \citenamefont
  {Schafer}}]{Kulander1996}%
  \BibitemOpen
  \bibfield  {author} {\bibinfo {author} {\bibfnamefont {K.~C.}\ \bibnamefont
  {Kulander}}, \bibinfo {author} {\bibfnamefont {F.~H.}\ \bibnamefont {Mies}},
  \ and\ \bibinfo {author} {\bibfnamefont {K.~J.}\ \bibnamefont {Schafer}},\
  }\href {\doibase 10.1103/PhysRevA.53.2562} {\bibfield  {journal} {\bibinfo
  {journal} {Phys. Rev. A}\ }\textbf {\bibinfo {volume} {53}},\ \bibinfo
  {pages} {2562} (\bibinfo {year} {1996})}\BibitemShut {NoStop}%
\bibitem [{\citenamefont {Majety}\ and\ \citenamefont
  {Scrinzi}(2015{\natexlab{a}})}]{Majety2015}%
  \BibitemOpen
  \bibfield  {author} {\bibinfo {author} {\bibfnamefont {V.~P.}\ \bibnamefont
  {Majety}}\ and\ \bibinfo {author} {\bibfnamefont {A.}~\bibnamefont
  {Scrinzi}},\ }\href {\doibase 10.1103/PhysRevLett.115.103002} {\bibfield
  {journal} {\bibinfo  {journal} {Phys. Rev. Lett.}\ }\textbf {\bibinfo
  {volume} {115}},\ \bibinfo {pages} {103002} (\bibinfo {year}
  {2015}{\natexlab{a}})}\BibitemShut {NoStop}%
\bibitem [{\citenamefont {Majety}\ \emph
  {et~al.}(2015{\natexlab{a}})\citenamefont {Majety}, \citenamefont
  {Zielinski},\ and\ \citenamefont {Scrinzi}}]{Majety2015e}%
  \BibitemOpen
  \bibfield  {author} {\bibinfo {author} {\bibfnamefont {V.~P.}\ \bibnamefont
  {Majety}}, \bibinfo {author} {\bibfnamefont {A.}~\bibnamefont {Zielinski}}, \
  and\ \bibinfo {author} {\bibfnamefont {A.}~\bibnamefont {Scrinzi}},\ }\href
  {\doibase 10.1088/0953-4075/48/2/025601} {\bibfield  {journal} {\bibinfo
  {journal} {J. Phys. B}\ }\textbf {\bibinfo {volume} {48}},\ \bibinfo {pages}
  {025601} (\bibinfo {year} {2015}{\natexlab{a}})}\BibitemShut {NoStop}%
\bibitem [{\citenamefont {Majety}\ and\ \citenamefont
  {Scrinzi}(2015{\natexlab{b}})}]{Majety2015d}%
  \BibitemOpen
  \bibfield  {author} {\bibinfo {author} {\bibfnamefont {V.}~\bibnamefont
  {Majety}}\ and\ \bibinfo {author} {\bibfnamefont {A.}~\bibnamefont
  {Scrinzi}},\ }\href {\doibase 10.3390/photonics2010093} {\bibfield  {journal}
  {\bibinfo  {journal} {Photonics}\ }\textbf {\bibinfo {volume} {2}},\ \bibinfo
  {pages} {93} (\bibinfo {year} {2015}{\natexlab{b}})}\BibitemShut {NoStop}%
\bibitem [{\citenamefont {Majety}\ \emph
  {et~al.}(2015{\natexlab{b}})\citenamefont {Majety}, \citenamefont
  {Zielinski},\ and\ \citenamefont {Scrinzi}}]{Majety2015c}%
  \BibitemOpen
  \bibfield  {author} {\bibinfo {author} {\bibfnamefont {V.~P.}\ \bibnamefont
  {Majety}}, \bibinfo {author} {\bibfnamefont {A.}~\bibnamefont {Zielinski}}, \
  and\ \bibinfo {author} {\bibfnamefont {A.}~\bibnamefont {Scrinzi}},\ }\href
  {\doibase 10.1088/1367-2630/17/6/063002} {\bibfield  {journal} {\bibinfo
  {journal} {New J. Phys.}\ }\textbf {\bibinfo {volume} {17}},\ \bibinfo
  {pages} {63002} (\bibinfo {year} {2015}{\natexlab{b}})}\BibitemShut {NoStop}%
\bibitem [{\citenamefont {Majety}\ and\ \citenamefont
  {Scrinzi}(2015{\natexlab{c}})}]{Majety2015g}%
  \BibitemOpen
  \bibfield  {author} {\bibinfo {author} {\bibfnamefont {V.~P.}\ \bibnamefont
  {Majety}}\ and\ \bibinfo {author} {\bibfnamefont {A.}~\bibnamefont
  {Scrinzi}},\ }\href {\doibase 10.1088/0953-4075/48/24/245603} {\bibfield
  {journal} {\bibinfo  {journal} {J. Phys. B}\ }\textbf {\bibinfo {volume}
  {48}},\ \bibinfo {pages} {245603} (\bibinfo {year}
  {2015}{\natexlab{c}})}\BibitemShut {NoStop}%
\bibitem [{\citenamefont {Hiskes}(1961)}]{Hiskes1961}%
  \BibitemOpen
  \bibfield  {author} {\bibinfo {author} {\bibfnamefont {J.~R.}\ \bibnamefont
  {Hiskes}},\ }\href {\doibase 10.1103/PhysRev.122.1207} {\bibfield  {journal}
  {\bibinfo  {journal} {Phys. Rev.}\ }\textbf {\bibinfo {volume} {122}},\
  \bibinfo {pages} {1207} (\bibinfo {year} {1961})}\BibitemShut {NoStop}%
\bibitem [{\citenamefont {Scrinzi}(2010)}]{Scrinzi2010}%
  \BibitemOpen
  \bibfield  {author} {\bibinfo {author} {\bibfnamefont {A.}~\bibnamefont
  {Scrinzi}},\ }\href {\doibase 10.1103/PhysRevA.81.053845} {\bibfield
  {journal} {\bibinfo  {journal} {Phys. Rev. A}\ }\textbf {\bibinfo {volume}
  {81}},\ \bibinfo {pages} {053845} (\bibinfo {year} {2010})}\BibitemShut
  {NoStop}%
\bibitem [{\citenamefont {McCurdy}\ \emph {et~al.}(2004)\citenamefont
  {McCurdy}, \citenamefont {Baertschy},\ and\ \citenamefont
  {Rescigno}}]{McCurdy2004}%
  \BibitemOpen
  \bibfield  {author} {\bibinfo {author} {\bibfnamefont {C.~W.}\ \bibnamefont
  {McCurdy}}, \bibinfo {author} {\bibfnamefont {M.}~\bibnamefont {Baertschy}},
  \ and\ \bibinfo {author} {\bibfnamefont {T.~N.}\ \bibnamefont {Rescigno}},\
  }\href {\doibase 10.1088/0953-4075/37/17/R01} {\bibfield  {journal} {\bibinfo
   {journal} {J. Phys. B}\ }\textbf {\bibinfo {volume} {37}},\ \bibinfo {pages}
  {R137} (\bibinfo {year} {2004})}\BibitemShut {NoStop}%
\bibitem [{\citenamefont {Chelkowski}\ \emph {et~al.}(1995)\citenamefont
  {Chelkowski}, \citenamefont {Zuo}, \citenamefont {Atabek},\ and\
  \citenamefont {Bandrauk}}]{Chelkowski1995}%
  \BibitemOpen
  \bibfield  {author} {\bibinfo {author} {\bibfnamefont {S.}~\bibnamefont
  {Chelkowski}}, \bibinfo {author} {\bibfnamefont {T.}~\bibnamefont {Zuo}},
  \bibinfo {author} {\bibfnamefont {O.}~\bibnamefont {Atabek}}, \ and\ \bibinfo
  {author} {\bibfnamefont {A.~D.}\ \bibnamefont {Bandrauk}},\ }\href {\doibase
  10.1103/PhysRevA.52.2977} {\bibfield  {journal} {\bibinfo  {journal} {Phys.
  Rev. A}\ }\textbf {\bibinfo {volume} {52}},\ \bibinfo {pages} {2977}
  (\bibinfo {year} {1995})}\BibitemShut {NoStop}%
\bibitem [{\citenamefont {Ch{\^{a}}teauneuf}\ \emph {et~al.}(1998)\citenamefont
  {Ch{\^{a}}teauneuf}, \citenamefont {Nguyen-Dang}, \citenamefont {Ouellet},\
  and\ \citenamefont {Atabek}}]{Chateauneuf1998}%
  \BibitemOpen
  \bibfield  {author} {\bibinfo {author} {\bibfnamefont {F.}~\bibnamefont
  {Ch{\^{a}}teauneuf}}, \bibinfo {author} {\bibfnamefont {T.~T.}\ \bibnamefont
  {Nguyen-Dang}}, \bibinfo {author} {\bibfnamefont {N.}~\bibnamefont
  {Ouellet}}, \ and\ \bibinfo {author} {\bibfnamefont {O.}~\bibnamefont
  {Atabek}},\ }\href {\doibase 10.1063/1.475800} {\bibfield  {journal}
  {\bibinfo  {journal} {J. Chem. Phys.}\ }\textbf {\bibinfo {volume} {108}},\
  \bibinfo {pages} {3974} (\bibinfo {year} {1998})}\BibitemShut {NoStop}%
\bibitem [{\citenamefont {Bressanini}\ \emph {et~al.}(1997)\citenamefont
  {Bressanini}, \citenamefont {Mella},\ and\ \citenamefont
  {Morosi}}]{Bressanini1997}%
  \BibitemOpen
  \bibfield  {author} {\bibinfo {author} {\bibfnamefont {D.}~\bibnamefont
  {Bressanini}}, \bibinfo {author} {\bibfnamefont {M.}~\bibnamefont {Mella}}, \
  and\ \bibinfo {author} {\bibfnamefont {G.}~\bibnamefont {Morosi}},\ }\href
  {\doibase 10.1016/S0009-2614(97)00571-X} {\bibfield  {journal} {\bibinfo
  {journal} {Chem. Phys. Lett.}\ }\textbf {\bibinfo {volume} {272}},\ \bibinfo
  {pages} {370} (\bibinfo {year} {1997})}\BibitemShut {NoStop}%
\bibitem [{\citenamefont {Schaad}\ and\ \citenamefont
  {Hicks}(1970)}]{Schaad1970}%
  \BibitemOpen
  \bibfield  {author} {\bibinfo {author} {\bibfnamefont {L.~J.}\ \bibnamefont
  {Schaad}}\ and\ \bibinfo {author} {\bibfnamefont {W.~V.}\ \bibnamefont
  {Hicks}},\ }\href {\doibase 10.1063/1.1674078} {\bibfield  {journal}
  {\bibinfo  {journal} {J. Chem. Phys.}\ }\textbf {\bibinfo {volume} {53}},\
  \bibinfo {pages} {851} (\bibinfo {year} {1970})}\BibitemShut {NoStop}%
\bibitem [{\citenamefont {Pavi{\v{c}}i{\'{c}}}\ \emph
  {et~al.}(2005)\citenamefont {Pavi{\v{c}}i{\'{c}}}, \citenamefont {Kiess},
  \citenamefont {H{\"{a}}nsch},\ and\ \citenamefont {Figger}}]{Pavicic2005}%
  \BibitemOpen
  \bibfield  {author} {\bibinfo {author} {\bibfnamefont {D.}~\bibnamefont
  {Pavi{\v{c}}i{\'{c}}}}, \bibinfo {author} {\bibfnamefont {A.}~\bibnamefont
  {Kiess}}, \bibinfo {author} {\bibfnamefont {T.~W.}\ \bibnamefont
  {H{\"{a}}nsch}}, \ and\ \bibinfo {author} {\bibfnamefont {H.}~\bibnamefont
  {Figger}},\ }\href {\doibase 10.1103/PhysRevLett.94.163002} {\bibfield
  {journal} {\bibinfo  {journal} {Phys. Rev. Lett.}\ }\textbf {\bibinfo
  {volume} {94}},\ \bibinfo {pages} {163002} (\bibinfo {year}
  {2005})}\BibitemShut {NoStop}%
\end{thebibliography}%
\end{document}